\documentclass[]{JHEP3}
\usepackage{amsmath,amsfonts,amssymb,amsthm,amstext,amscd,eucal}

\newcommand{\be}{\begin{equation}}
\newcommand{\ee}{\end{equation}}
\newcommand{\bea}{\begin{eqnarray}}
\newcommand{\eea}{\end{eqnarray}}
\newcommand{\eO}{\mathcal{O}}
\newcommand{\eF}{\mathcal{F}}
\newcommand{\eP}{\cal{P}}

\newcommand{\bbibitem}[1]{\bibitem{#1}} 
\newcommand{\mc}{\mathcal}
\newcommand{\mbb}{\mathbb}
\newcommand{\K}{\mc{K}}

\def\nobbibitem{\let\bbibitem=\bibitem}

\def\t{{\tau}}
\def\t1c{{\tau_1^c}}
\def\g{{\gamma}}

\def\Mp{M_{P}}

\def\Mpl{M_{pl}}

\title{Topology from Cosmology}

\author{Vijay Balasubramanian$^{a,b}$, Per Berglund$^{c}$, Raul Jimenez$^{d,e}$, Joan Simon$^{f}$, Licia Verde$^{d,e}$ \\

$^a$ David Rittenhouse Lab., University of Pennsylvania, Philadelphia, PA 19104, USA \\
$^b$ School of Natural Sciences, Institute for Advanced Study, Princeton, NJ 08540, USA\\
$^c$ \it Department of Physics, University of New Hampshire, Durham, NH 03824, USA\\
$^d$ Institute of Space Sciences (CSIC-IEEC)/ICREA, UAB, Barcelona 08193, Spain\\
$^e$ Dept. of Astrophysical Sciences, Princeton University, Princeton NJ 08544, USA\\
$^f$ School of Mathematics and Maxwell Institute for Mathematical Sciences, King's Buildings, Edinburgh EH9 3JZ, Scotland\\

Email: \email{vijay@physics.upenn.edu, per.berglund@unh.edu, raulj@astro.princeton.edu, J.Simon@ed.ac.uk, lverde@astro.princeton.edu}}

\abstract{We show that cosmological observables can constrain the topology of the compact additional dimensions predicted by string theory.    To do this, we develop a general strategy for relating cosmological observables to the microscopic parameters of the potentials and field-dependent kinetic terms of the multiple scalar fields that  arise in the low-energy limit of string theory.  We apply this formalism to the Large Volume Scenarios in Type IIB flux compactifications where analytical calculations are possible.   
Our methods generalize to other settings.
}

\begin{document}






\section{Introduction}
\label{sec:intro}

Most string theories achieve unification of the forces by involving six dimensions of spacetime beyond the familiar four.    A realistic model, with reduced symmetries and degrees of freedom at low energies, typically involves taking the extra dimensions to be a compact Ricci-flat manifold, a so called Calabi-Yau manifold \cite{Candelas:1985en}.    The topological data of this manifold, encoded in its cohomology classes, enter into low-energy physics in terms of particle spectra and the structure of potentials for scalar fields (``moduli'') that encode the sizes and shapes of surfaces within the six hidden dimensions.

In a scenario where inflation is realized by rolling moduli scalar fields\footnote{For initial explorations in this subject, see \cite{old}. For some reviews on string cosmology, see \cite{reviews}. For references with a similar philosophy to the one advocated in this work, see \cite{similar}.} , it is possible that the microscopic information, such as the topological data, entering the shape of the potential, will affect cosmological observables, in particular the Cosmic Microwave Background (CMB) fluctuation spectrum.   In this paper we show that cosmological observables can be surprisingly sensitive to the microscopic parameters controlling the scalar potential.  For example, in a simple case with two scalar fields, we show that the minimal variation in the Euler number of the six hidden dimensions can destroy slow roll inflation, unless specifically compensated elsewhere in the model.  Thus, precision cosmological measurements could  limit the topological invariants of extra dimensions -- this would strongly constrain string theoretic models of unification. The bulk of this paper studies the Large Volume Scenarios \cite{LVS} in which analytical calculations are possible, but we expect that similar studies can be carried out in other inflationary settings arising from string theory.

The main technical hurdle we have to cross is  the indirect connection between cosmological observables and the microscopic parameters of a string compactification.    Any string theoretic model requires  certain {\it a priori} choices -- a corner in the space of string theories (Type IIB, heterotic, M-theory etc.) and a choice of compactification to four dimensions.   Given these choices the low-energy field content (coming from light closed strings and, if D-branes are present, light open strings) is fixed.  Typically, in order to have theoretical control over the low-energy effective field theory, the string coupling is assumed to be weak and the compactification is taken as a six-dimensional manifold with all length scales much larger than the string length.  In addition, various fields may have background expectation values.     The low-energy Lagrangian is then determined in terms of microscopic parameters $\{\alpha_m\}$ (including the string coupling, string length, and topological parameters of the hidden dimensions), typically as a power series expansion that is determined order by order in appropriate small quantities.  We make these assumptions, knowing that  the real world could lie in a different part of the configuration space where the coupling is strong or where the compactification to four dimensions does not even involve a conventional geometry.
   
In our setting, inflation arises from the rolling of scalar fields $\{\phi^a(t)\}$ with a Lagrangian determined by the parameters $\{\alpha_m\}$.    Cosmological observables arising from inflation are usually reported at a fixed wavenumber $k$ that crossed the horizon about $13$  e-foldings before the end of inflation \cite{liddle-lyth}, during an epoch when the fields took values $\{ \phi^a(t_0) \}$, where $t_0$ is the time at which the fixed wavenumber crossed the horizon.   Thus we will find that the scalar potential will be most conveniently parametrized in terms of a set of effective parameters $z_i(\{ \alpha_m \}, \{ \phi^a(t_0) \})$.    The inflationary observables $ \{\eO_r \}=\{n_s,\,\alpha_s,\,r,\,{\cal P}\,,\dots\}$ (where $n_s$ stands
for the spectral index, $\alpha_s$ for its running, $r$ for the ratio of the tensor to scalar perturbations and ${\cal P}$ for the
power spectrum of the adiabatic scalar perturbations) will depend on these effective parameters:
\begin{equation}
  \eO_r =  \eO_r(\{z_i\}) = \eO_r(\{ \phi^a (t_0) \},\,\{\alpha_m\})\, .
 \label{obs-th}
\end{equation}
Because  there will generally be fewer observables than parameters, and because only the combinations $z_i$ will appear in $\eO_r$, cosmological measurements can only limit $\{\alpha_m\}$ and $\{\phi^a(t_0)\}$ to jointly lie on some (possibly infinite) surface. These degeneracy surfaces
are fully determined by the values of the parameters $z_i$ matching current observations, but the
addition of further observables from particle physics and cosmology might break some of this degeneracy.  Because topological parameters are discrete and sparse, even such weak limits can be very constraining on string theoretic models.  

We wish to explore how bounds on the cosmological observables $\eO_r$ can  be translated into  constraints on the microscopic parameters $\{ \alpha_m \}$.  In the spirit of the Fisher matrix approach used in cosmology,  we  will linearize the equations in a small region of parameter space around a good solution (i.e. a set of microscopic parameters and initial conditions that satisfies slow roll conditions, realizes the correct number of e-foldings  and that matches the best fit cosmological parameters). To do so, we will consider  small changes in the microscopic parameters and quantify the corresponding changes in the observables. This will then relate  bounds on observable quantities to constraints on the microscopic parameters.  Specifically, an uncertainty $\delta {\cal O}_r$ in the observables will be mapped into an uncertainty in the microscopic parameters:
\begin{equation}
\delta \eO_r \equiv \vec{v}\cdot\vec{\nabla} \eO_r (\{\phi^a(t_0)\},\,\{\alpha_m\})\,,
\label{sensitivity1}
\end{equation}
where $\vec{v}$ is a vector of variations in the space of parameters. Given that we will work in the slow roll approximation we will also have to be careful that the variations in parameters preserve the slow roll conditions.  Geometrically, this is achieved by restricting the set of variations $\vec{v}$ to those preserving the set of constraints $\eF_A (\alpha_m,\,\phi^a(t_0))$:
\begin{equation}
  \vec{v} \cdot \vec{\nabla} \eF_A = 0 \quad \forall\,\, A
\end{equation}
Once we work with the proper subspace of variations $\vec{v}$, we can study
the sensitivity of the observables under their action, and compare these
with the current observational bounds on each of our observables as a function
of the point in parameter space and the initial conditions.

In section 2 we set up the general framework  of ${\cal N}=1$ supergravity in four dimensions and the Large Volume Scenario that we use. We then derive the dynamical regime in which slow-roll inflation holds in our models in  section 3, and discuss how the multi-field scenario naturally generalizes the standard single-field slow-roll parameters.  In section 4 the cosmological observables are expressed in terms of the generalized slow-roll parameters. This allows us to solve for the microscopic parameters in terms of the ratio of the tensor to scalar perturbations, $r$ and the spectral index, $n_s$. 
In section 5, we consider the sensitivity of the observables to variations of the microscopic parameters and we summarise our main results in a discussion in
section 6. The analytical properties of the scalar potential and its derivatives, and the computation of the different observables in our models are presented in detail in the final appendices.

\section{String theory set-up}

We will examine the rolling of scalar fields in an effective supergravity theory, since the  low energy Lagrangian of any supersymmetric string theory can be written in this language, even though supersymmetry will be broken in the ground state of a realistic model.    Much of the notation and technique is thereby applicable to any setting that has a valid description in supergravity, although the particular mechanism for realizing inflation, and the detailed appearance of topological data in the scalar potential, will be specific to our example.

An $\mc{N} = 1$ supergravity in four dimensions is completely specified by a K\"ahler potential $(\K)$ and superpotential $(\hat{W})$. 
The K\"{a}hler potential is a real function of the complex scalar fields and so will only depend on $\phi_i - \bar{\phi}_{\bar{i}}$, while the superpotential is a holomorphic function depending only on $\phi_i$ and not $\bar{\phi}_{{\bar i}}$.
Focusing on the  dynamics of the moduli scalar fields relevant for inflation, the action is (we will work in the Einstein frame; see the Appendix of \cite{CQS} for conventions)
\begin{equation}
S_{\mc{N}=1} = \int d^4 x \sqrt{-g} \left[ \frac{M_P^2}{2} \,R - {\cal G}{i\bar{j}} D_\mu \phi^{i} D^\mu \bar{\phi}^j
- V(\phi_i, \bar{\phi}_i) \right]\,.
 \label{eq:action}
\end{equation}
The scalar potential is given by 
\begin{eqnarray}
V(\phi_i, \bar{\phi}_i)  &=& e^{\K/M_P^2} \left({\cal G}^{i \bar{j}} D_i \hat{W} D_{\bar{j}} \bar{\hat{W}} - \frac{3}{M_P^2} \hat{W}
\bar{\hat{W}} \right) + 
V_{\text{uplift}} \label{eq:potential}
\\
D_i \hat{W} &=& \partial_i \hat{W} + \hat{W} \partial_i \K  \\
{\cal G}_{i\bar{j}} &=& \partial_i \partial_{\bar j} \K
\end{eqnarray}
The derivatives $\partial_i$ and $\partial_{\bar{i}}$ differentiate with respect to the $\phi_i$ and $\bar{\phi}_{\bar{i}}$ dependence.   Here $M_P$ is  the reduced Planck mass in four dimensions $M_P = \frac{1}{(8 \pi G)^{1/2}} = 2.4 \times 10^{18} \, \textrm{GeV}$.   The specific form of $\K$ and $\hat{W}$ depends on the particular choice of string theory and compactification, and this is where the microscopic parameters of string theory enter the effective Lagrangian.  The term $V_{\text{uplift}}$  will  include the effects of supersymmetry breaking arising from other sectors of the theory.

\subsection{Type IIB flux compactifications}
\label{fluxsec}
We will demonstrate our methods in the context of Type IIB string theory compactified to four dimensions on a Calabi-Yau orientifold because the scalar potential in this case is well-understood and realistic four-dimensional models can be constructed \cite{GKP,KKLT,DouglasDenef,LVS,CQS,Allenach}.  Readers wishing to skip the technical details of the underlying string theory can move on to the next section where the effective potential relevant to inflation is reported.

We will work in the four dimensional Einstein frame and follow the conventions of \cite{CQS}.   Then, after including the leading perturbative and non-perturbative corrections of string theory, the
K\"ahler potential and superpotential are
\begin{eqnarray}
\frac{\mc{K}}{M_P^2} & = & - 2 \ln \left(\mc{V} + \frac{ \xi \, g_s^{\frac{3}{2}}}{2 e^{\frac{3 \phi}{2}}} \right)
- \ln(-i(\tau - \bar{\tau})) - \ln \left(-i \int_{M} \Omega \wedge \bar{\Omega}\right), \nonumber \\
\hat{W}  & = & \frac{g_s^{\frac{3}{2}} M_P^3}{\sqrt{4 \pi} l_s^2} \left( \int_{M}
 G_3 \wedge \Omega + \sum A_i e^{-a_i T_i} \right)
\equiv \frac{g_s^{\frac{3}{2}} M_P^3}{\sqrt{4 \pi}} W.
\label{eq:potentials}
\end{eqnarray}
Here $g_s$ is the string coupling, $l_s$ is the string length,  $\Omega$ is the holomorphic three-form on the Calabi-Yau manifold $M$,  $G_3$ is the background field (flux) that is chosen to thread 3-cycles in $M$ and
\be
\xi = -\frac{\zeta(3) \, \chi(M)}{2 (2 \pi)^3}
\label{xidef}
\ee
where $\chi$ is Euler number of $M$.  The axion-dilaton field is $\tau=C_0+ i \, e^{-\phi}$, and the integrals involving $\Omega$ are implicitly functions of the complex structure moduli.  The fields  $T_i = \tau_i + i b_i$ are the complexified K\"ahler moduli where $\tau_i$ is a  4-cycle volume (of the divisor $D_i\in H_4(M,\mbb{Z})$)  and $b_i$ is its axionic partner arising ultimately from the 4-form field.    Here $a_i = 2\pi/N_i$ for some integer $N_i$, for each field, that is determined by the dynamical origin of the exponentials in the superpotential ($N_i = 1$ for brane instanton contributions, $N_i > 1$ for gaugino condensates). Finally, $\mc{V}$ is the dimensionless classical volume of the compactification manifold $M$ (in Einstein frame, but measured in units of the string length).   In terms of the K\"ahler class $J=\sum_i t^i D_i$ (by Poincar\'{e} duality $D_i\in H^2(M,\mbb{Z}$)), with the $t^i$ measuring the areas of 2-cycles, $C_i$,
\begin{equation}
{\mc V} =  \int_M J^3 = \frac{1}{6} \kappa_{ijk} t^i t^j t^k~,
\label{CYvolume}
\end{equation}
where $\kappa_{ijk}$ are the intersection numbers of the manifold.   ${\mc V}$ should be understood as an implicit function of the complexified 4-cycle moduli $T_k$ via the relation
\begin{equation}
\tau_i = \partial_{t_i} {\mc V} = \frac{1}{2} \kappa_{ijk} t^j t^k~.
\label{4to2cycles}
\end{equation}
The factors of $g_s$ in the above follow the conventions of \cite{CQS}. 
In particular, when evaluated in the vacuum, the string and Einstein frames are identical.
 Note that the K\"ahler potential $\K$ has mass dimension 2 whereas the superpotential $\hat{W}$ has mass dimension 3.    In these conventions, the reduced Planck mass in four dimensions, $M_P$,  is related to the string scale $(l_s)$  by
\begin{equation}
M_P^2 = \frac{4 \pi \mc{V}_0}{g_s^2 l_s^2} \quad \textrm{ or } \quad
m_s = \frac{g_s}{\sqrt{4 \pi \mc{V}_0}} M_P\,,
\label{eq:mplanck}
\end{equation}
where $\mc{V}_0 = \langle \mc{V} \rangle$ is the  volume of the compactification manifold $M$ at the minimum of the scalar potential.

There are additional perturbative corrections to $\mc{K}$ in (\ref{eq:potentials}), but we have kept the terms that give the leading contributions to the scalar potential in the large $\mc{V}$ limit of interest to us \cite{bergetal}.  In particular, expanding $\mc{K}$ to linear order in $\xi$ gives a consistent approximation in inverse powers of $\mc{V}$.   We have also assumed that all of the K\"ahler moduli $T_i$ appear in the superpotential (see \cite{DouglasDenef} for examples) and that we use a basis of 4-cycles such that the exponential terms in $\hat{W}$ take the form $exp(- a_i \, T_i)$.   As these exponentials arise from an instanton expansion, in order to only keep the first term as we have done, the 4-cycle volumes must be sufficiently large to ensure that $a_i T_i \gg 1$. 

 Finally, the form of the term $V_{\text{uplift}}$ in (\ref{eq:potential}) depends on the kind of supersymmetry breaking effects that arise from other sectors of the theory.  
We take
\begin{equation}
  V_{\text{uplift}} = \frac{\gamma}{\mc{V}^2}\,
  \label{uplift}
\end{equation}
which will describe the energy of a space-filling antibrane \cite{KKLT}, fluxes of gauge fields living on D7-branes \cite{BKQ}, or the F-term due to a non-supersymmetric solution for the complex structure/axion-dilaton moduli \cite{Saltman:2004sn}.

It was shown in \cite{KKLT} that a generic choice of background fields $G_3$ causes all the complex structure moduli and the axion-dilaton to acquire string scale masses without breaking supersymmetry.  They are then decoupled from the low-energy theory and their contributions to $\mc{K}$ and $\hat{W}$ are constants for our purposes\footnote{In the case of the F-term breaking due to the complex structure/axion-dilaton moduli \cite{Saltman:2004sn}, the contribution of the complex structure and axion-dilaton moduli to the scalar potential does depend on the volume \eqref{uplift}.}:
\begin{eqnarray}
\mc{K} & = & - 2 M_P^2 \, \ln \left(\mc{V} + \frac{ \xi}{2 } \right)
-\ln \left( {2\over g_s}\right) + \mc{K}_0 , \nonumber \\
\hat{W}  & = & \frac{g_s^{\frac{3}{2}} M_P^3}{\sqrt{4 \pi} } \left( W_0 + \sum_i A_i e^{-a_i T_i} \right)  \equiv \frac{g_s^{\frac{3}{2}} M_P^3}{\sqrt{4 \pi}} W \, .
\label{eq:potentials2}
\end{eqnarray}
It was shown in \cite{LVS} that, when the Euler number, $\chi < 0$,  for generic values of $W_0$ (and hence of the background fluxes $G_3$), the scalar potential for the K\"ahler moduli has a minimum where the volume ${\mc V}$  of the Calabi-Yau manifold $M$ is very large -- the associated energy scale is a few orders of magnitude lower than the GUT scale.     Furthermore, in these Large Volume Scenarios  there is a natural hierarchy -- one of the K\"ahler moduli is much larger than the others and dominates the volume of the manifold.  Because the Large Volume Scenarios exist for generic choices of fluxes (unlike the KKLT scenarios \cite{KKLT} which require fine-tuning) they are the statistically favored setting for Type IIB model building.   For our purposes they are also attractive because the scalar potential admits an expansion in inverse powers of the large volume $\mc{V}$.  This will allow us to carry out analytical calculations of inflation arising from K\"ahler moduli rolling towards the large volume minimum of the potential.

Several previous works have considered inflation in the large volume setting, e.g., \cite{CQ,roulette,ourprevious,jimmy}.  Here we  focus on whether the resulting cosmological observables can meaningfully constraint the microscopic parameters appearing in (\ref{eq:potentials2}), i.e., $g_s$, $l_s$, and topological quantitites such as the  intersection numbers $\kappa_{ijk}$ and the Euler number $\chi$.    In our analyses the uplift parameter $\gamma$ is freely tunable, but in order for the scalar potential to have a minimum at finite volume for $M$, it will be necessary for $\gamma \sim 1/\mc{V}$ \cite{CQS}.   For our purposes we will also treat the coefficients $A_i$ as independent parameters although they are in fact dynamically determined once the compactification manifold $M$ is specified.  
Thus, this kind of analysis offers a consistency check, i.e. the parameters  that come out of the analysis
need to correspond to an actual compactification manifold.

\subsection{Two field models in the Large Volume Scenario}

We are interested in finding a slow roll scenario for inflation in which we can test the sensitivity of the observables to the microscopic parameters.     Slow roll inflation can occur in  a region of the field space where the potential is positive and very flat.  We will look for this in the Large Volume Scenarios described above, where, at the minimum of the scalar potential, there is a hierarchy amongst the K\"ahler moduli
\begin{equation}
\tau_1 \gg \tau_2, \tau_{3} , \tau_4 \cdots
\label{modulihierarchy}
\end{equation}
which we will use to simplify the effective potential.   For simplicity, we work with two complex fields $\{T_1,\,T_2\}$, a situation that can arise either if the compactification manifold $M$ has only two K\"ahler moduli, or if the remaining fields play a spectator role.  However, all our expressions easily generalize to the many-field setting, as we will see shortly. 
Also for transparency of the equations, we will assume  that the intersection numbers $k_{ijk}$ are such that in the basis of 4-cycles, $\tau_i$, the volume takes the diagonal form, 
\be
{\mc V} =\mu \tau_1^{3/2}-\lambda \tau_2^{3/2} \equiv \frac{1}{x}\,\left(1-\lambda\,x\,\tau_2^{3/2}\right)\,,\quad x^{-1}\equiv \mu \tau_1^{3\over2}\,,
\label{largevolume}
\ee
where $1/x$ is the dominant contribution to the volume of the compactification manifold.  Thus, $\lambda$ and $\mu$ are the only non-vanishing intersection numbers in our compactification manifold.
This is also a restriction that is easily relaxed.
Including the leading order $\alpha'$ correction we have
\be
{\mc V} + {\xi\over 2} = x^{-1}\left(1+\beta_1-\beta_2\right)\,,\quad\beta_1\equiv x{\xi\over 2}\,,\quad \beta_2\equiv \lambda\,x\,\tau_2^{3/2}
\ee
  In the Large Volume Scenarios, at the minimum of the scalar potential
\be
x \ll 1\, ,
\label{largevolume2}
\ee 
and so we can organize the scalar potential  \eqref{eq:potential}, which now depends on four real scalar fields $\{\tau_1,\,\tau_2,\,b_1,\,b_2\}$, in powers of $x$.  We will take the axionic fields $\{b_i\}$ to be stabilized at their minima, and consider using either $\tau_1$ or $\tau_2$ as the inflaton.    See \cite{racetrack,roulette} for discussions of inflation produced by rolling in other directions in the scalar field space.

Since $\tau_1 \gg \tau_2 \gg 1$ in the Large Volume Scenario where (\ref{eq:potentials2}) is valid, we will ignore all the exponentials in $\tau_1$ (which are completely negligible) and define
\be
y\equiv a_2\tau_2 \, .
\label{ydefinition}
\ee
Furthermore, we will work in the analytical regime defined by
\be
  \beta_1 \ll 1\,, \quad \beta_2 \ll1\,, \quad \frac{A_2\,e^{-y}}{|W_0|} \ll1\,.
 \label{eq:approx}
\ee
In addition to the large volume assumption, the first condition requires that the ratio of the intersection numbers $\lambda/\mu$ should not be too large and the second condition requires the same of the Euler number (to which $\xi$ is proportional (\ref{xidef})).   These topological conditions are natural since these numbers tend to be of order a few hundred  in typical examples.  The third condition requires that $A_2/|W_0|$ not be too large.    In the KKLT scenarios \cite{KKLT}, $A_2$ is taken to be $\eO(1)$ and $W_0$ is fine tuned to be very small.   However, in the Large Volume Scenarios \cite{LVS}, no such fine tuning of $W_0$ is needed, and for generic fluxes this condition will be satisfied also.

Subject to these conditions, the scalar potential (\ref{eq:potential}) can be approximated by
\begin{eqnarray}
   V_{\text{uplift}} &=& \gamma\,x^2\left(1 + \eO(\beta_2)\right)\,. \nonumber \\
   V_2 &=& -\frac{g_s^4\,M_P^4\,|W_0|^2}{4\pi}\,2 y\,\frac{A_2 \,e^{-y}}{|W_0|}\,x^2 
  \left(1 + \eO\left(\ell_i\right)\right) \nonumber \\
   V_3 &=& \frac{g_s^4\,M_P^4\,|W_0|^2}{4\pi}\,\frac{4 y^2}{3}\,\left(\frac{A_2 \,e^{-y}}{|W_0|}\right)^2\,\frac{x^2}{\beta_2}\left(1 +
   \eO\left(\ell_i\right)\right) \nonumber \\
   V_4 &=& \frac{g_s^4\,M_P^4\,|W_0|^2}{4\pi}\,\frac{3\,(\xi\,x)}{8}\,x^2\left(1 + \eO\left(\ell_i\right) \right)\,
 \label{eq:potapprox}
\end{eqnarray}
where 
the $\ell_i$ are any of the small quantities in (\ref{eq:approx}).  Recalling from Sec.~\ref{fluxsec} that  near the minimum of the scalar potential   $\gamma \sim 1/\mc{V} \sim x$ and $y_2 \sim \ln \mc{V}$, we see that in the vicinity of this minimum all the terms of the potential are of $\eO(x^3)$. 

It will be convenient to introduce the notation:
\begin{equation}
V = M_P^4 \hat\g x_0^2\Big(\frac{x^2}{x_0^2}  + 3  z_3\,\frac{x^3}{x_0^3}
+ 4  z_2\,\frac{\sqrt{y}}{\sqrt{y_0}}\, \frac{x}{x_0}\, e^{2(y_0-y)} -
2z_1\, \frac{x^2}{x_0^2}\,\frac{y}{y_0}\, e^{y_0-y}\Big)\,,
\label{tecpot}
\end{equation}
where we have defined $\hat \gamma \equiv \gamma/M_P^4$
and 
the three parameters $\{z_1,\,z_2,\,z_3\}$ as
\begin{eqnarray}
  z_1 &=& \frac{g_s^4\,|W_0|^2}{4\pi\hat\g}\,y_0\,\frac{A_2\,e^{-y_0}}{|W_0|}\,, \label{z1} \\
  z_2 &=& \frac{g_s^4\,|W_0|^2}{4\pi\hat\g}\,\frac{y_0^2}{3\beta_0}\,\left(\frac{A_2\,e^{-y_0}}{|W_0|}\right)^2\,, \label{z2} \\
  z_3 &=& \frac{g_s^4\,|W_0|^2}{4\pi\hat\g}\,\frac{1}{8}\,\xi\,x_0\,. \label{z3}
\end{eqnarray}
with 
\be
\beta_0 = \beta_2|_{x=x_0,y=y_0} \, .
\label{beta0def}
\ee
The $z_i$ depend on the microscopic parameters of the model as well as the {\it initial conditions} $\{x_0,\,y_0\}$.   (Strictly speaking we will take $x_0,y_0$ to correspond to field values about 13 e-foldings before the end of inflation, since the inflationary observables are evaluated at wavenumbers that were freezing out during that epoch.   The total number of e-foldings should be about $50-60$ to achieve the observed degree of homogeneity in the universe -- our ``initial'' conditions, therefore describe the field values at a particular point in the inflationary evolution \cite{liddle-lyth}.) In sum, while our model is specified by a certain number of microscopic parameters ($\mu,\lambda,\xi, g_s, l_s, a_2, A_2$), the dynamics of the inflaton will be controlled by the $z_i$ which include the choice of initial conditions.   Cosmological observables resulting from this model of inflation will constrain the $z_i$ to lie within some bounds, and this could be achieved by many degenerate choices of microscopic parameters and the initial conditions.

\paragraph{Generalization to many fields: }  

The above form of the potential is easily generalized when there are $n$ K\"ahler moduli, with  
$n>2$. We assume that the volume takes the diagonal form, including the leading order $\alpha'$ correction,
\be
{\mc V}+{\xi\over 2} =\mu \tau_1^{3/2}+\beta_1-\sum_{i=2}^n\lambda_i \tau_i^{3/2} \equiv \frac{1}{x}\,\left(1+\beta_1-\sum_{i=2}^n \beta_i\right) ~,\quad \beta_i\equiv\lambda_i x \tau_i^{3/2}~.
\label{volumemulti}
\ee
As in the previous analysis, the scalar potential is then given by (here $y^i=a_i\tau_i$)
\begin{equation}
V = M_P^4 \hat\g x_0^2\left(\frac{x^2}{x_0^2}  + 3  z_3\,\frac{x^3}{x_0^3}
+ \sum_{i=2}^n\left(4  z^i_2\,\frac{\sqrt{y^i}}{\sqrt{y^i_0}}\, \frac{x}{x_0}\, e^{2(y^i_0-y^i)} -
2z^i_1\, \frac{x^2}{x_0^2}\,\frac{y^i}{y^i_0}\, e^{y^i_0-y^i}\right)\right)\,,
\label{tecpotmany}
\end{equation}
with the parameters $\{z^i_1,\,z^i_2\}$ as (with $z_3$ given by~(\ref{z3}))
\begin{eqnarray}
  z^i_1 &=& \frac{g_s^4\,|W_0|^2}{4\pi\hat\g}\,y^i_0\,\frac{A\,e^{-y^i_0}}{|W_0|}\,, \label{z1i} \\
  z^i_2 &=& \frac{g_s^4\,|W_0|^2}{4\pi\hat\g}\,\frac{\left(y^i_0\right)^2}{3\beta_{0i}}\,\left(\frac{A\,e^{-y^i_0}}{|W_0|}\right)^2\,, \label{z2i} 
\end{eqnarray}
with $\beta_{0i}=\beta_i|_{x=x_0,y^i=y^i_0}$.

\subsection{Diagonalizing the Metric}

The kinetic energy is given to leading order in $\beta_i$ by: 
\be
K = {\Mpl^2\over 2} \left( {2\over 3}(1-{5\over 4}\beta_1 +{5\over 2}\beta_2)\left({\dot x\over x}\right)^2 + 3\beta_2 {\dot x\over x}{\dot y\over y} + {3 \beta_2\over 4} \left({\dot y\over y}\right)^2\right)
\ee
where all dots stand for time derivatives $d/dt$. 
\footnote{To compute connection coefficients and higher order derivatives of the metric we will need the metric to higher order in $\beta_i$ as reported in Appendix A.}  
In the Large Volume Scenario it is possible to diagonalize the metric on the field space, up to further corrections of ${\cal O}(\beta_2)$. We write the 
kinetic energy as 
\be
K = {\Mpl^2 \over 2}\left({3\over 2}\left(\dot\tau_1\over\tau_1\right)^2\left(1-{5\over 4}\beta_1 -2\beta_2\right) +{3\over 4}\beta_2\left({\dot\tau_2\over\tau_2}-3 {\dot\tau_1\over\tau_1}\right)^2\right)~.
\label{taumetric}
\ee
Since $\beta_i<<1$, we can approximate the coefficient $\left(1-{5\over 4}\beta_1 -2\beta_2\right)\approx 1$ and by introducing the new fields
\be
q^1 =\Mpl \sqrt{3\over2}\log{\tau_1}\,,\quad q^2 =\Mpl{2\over\sqrt{3}} \sqrt{\lambda\over \mu}\left(\tau_2\tau_1^{-3}\right)^{3\over4}~,
\ee 
we get
\be
K = {1\over 2} \left(\left(\dot q^1\right)^2+ \,e^{\sqrt{6}\,q^1/\Mpl} \left(\dot q^2\right)^2\right) =
 {1\over 2} \left(\left(\dot q^1\right)^2+ \Mpl^2{4\over 3}\beta_2 \left({\dot q^2\over q^2}\right)^2\right)
 \label{kineticdiagonal}
\ee
In the case of more than two fields, assuming that the volume takes the diagonal form given by~(\ref{volumemulti}), the kinetic energy can similarly be approximately diagonalized, where we drop terms of ${\cal O}(\beta_i)$ in expressions of the form $1+{\cal O}(\beta_i)$. We get 
\be
K = {1\over 2} \left(\left(\dot q^1\right)^2+ e^{\sqrt{6} q^1} \sum_{i=2}\left(\dot q^i\right)^2\right) =
 {1\over 2} \left(\left(\dot q^1\right)^2+ \Mpl^2 {4\over 3} \sum_{i=2}\beta_i \left({\dot q^i\over q^i}\right)^2\right)
 \label{diagonalkinetic}
\ee
where the new fields are
\be
q^1 =\Mpl \sqrt{3\over2}\log{\tau_1}\,,\quad q^i =\Mpl{2\over\sqrt{3}}  \sqrt{\lambda_i\over \mu}\left(\tau_i\tau_1^{-3}\right)^{3\over4}~,
\label{diagonalfieldbasis}
\ee 

\section{A slow roll scenario}

There are two main challenges in analyzing the dynamical conditions in our scenarios under which slow roll inflation can take place: (a) there are many fields involved in the motion, and (b) since the metric in field space is not flat, kinetic terms $(1/2)\,G_{ij} \dot{\tau}^i \dot{\tau}^j$ are not canonically normalised.   In this section we will extract the general slow roll conditions for multi-field inflation with non-canonical kinetic terms and then apply the analysis to our Large Volume Scenarios.

Using the standard convention that repeated indices are summed over and writing
\begin{equation*}
  K = \frac{1}{2}\,G_{ij}\dot{q}^i\,\dot{q}^j={1\over 2} \dot q^i \dot q_i \,,\,\,  {\rm where}\,\, 
  \dot{q}_i = G_{ij} \, \dot{q}^j\,,\,\,
  {\rm and} \,\,
  \Gamma^i_{jk}={1\over 2} G^{il}\left({\partial G_{lk}\over \partial q^j} +
  {\partial G_{lj}\over \partial q^k} - {\partial G_{jk} \over \partial q^l}\right)
  \, ,
\end{equation*}
for the kinetic energy $K$ and Christoffel connection $\Gamma^i_{jk}$, 
the classical equations of motion for the above scalar fields, $q^i$, coupled to gravity are given by
\begin{eqnarray}
  H^2 &=& \frac{1}{3\,\Mpl^2}\left(K +
  V\right)\,,
 \label{eq:einstein}\\
  \ddot{q}^i  + 3H\dot{q}^i + \Gamma^i_{jk} \dot{q}^j\dot{q}^k& = & - G^{ij}\,\frac{\partial V}{\partial q^j}\,. 
  \label{eq:motion1} 
  \end{eqnarray}
  It is convenient to rewrite the acceleration and connection terms by using a covariant definition when defining derivation with respect to $t$
  where we use the covariant derivative when defining derivation with respect to $t$ \cite{stewart} , ie
  \begin{equation}
   \frac{d}{dt} = \dot q^i \nabla_i ~~~~\Longrightarrow~~~~
   \frac{d^2q^i}{dt^2} \equiv  \ddot{q}^i +  \Gamma^i_{jk} \dot{q}^j\dot{q}^k~.
  \ee
We will generally use this covariant definition of the acceleration, which allows us to write the equations of motion~(\ref{eq:motion1})  as
\be
\frac{d^2q^i}{dt^2} + 3H\dot{q}^i = - G^{ij}\,\frac{\partial V}{\partial q^j}\,. 
\label{eq:motion1-covariant}
\ee
The slow roll approximation requires:
\begin{itemize}
\item[(i)] The energy density to be dominated by the potential energy density: 
\begin{equation}
  K \ll V \quad \Rightarrow \quad H^2 \approx \frac{V}{3\,\Mpl^2}\,, 
  \label{slowroll1}
\end{equation}
which implies that
\be
 \hat \epsilon \equiv \frac{K}{\Mpl^2 \, H^2}=\frac{\dot{q}^i\,\dot{q}_i}{2\,\Mpl^2\,H^2} \ll 1
\label{epsilonhat}
\ee
The parameter $\hat\epsilon$ defined here is the first slow roll parameter, and is the multi-field analogue of the conventional single-field first slow roll parameter.
\item[(ii)] The friction term in the scalar field classical equations of motion dominates over the (covariant) acceleration term :
\begin{equation}
   \left|\frac{d^2q^i}{dt^2}\right| \ll |3H\dot q^i| \quad \Rightarrow \quad 
   3H\dot{q}^i \approx - G^{ij}\, \frac{\partial V}{\partial q^j}\quad\forall\, i\,.
 \label{sloweom}
\end{equation}
\end{itemize}
Note that this gives as many conditions as the number of fields, $q^i$.  Thus
by solving for $\dot q^i$ from~(\ref{sloweom}) and then differentiating 
 with respect to time gives a consistency condition for the validity of the slow roll approximation~\footnote{We assume that the term proportional to $\frac{d}{dt}\left({1\over 3H}\frac{d^2q^i}{dt^2}\right)$ can be neglected compared to the acceleration term.}:
\begin{equation}
\left|\frac{d^2q^i}{dt^2}\right| \approx \left|\dot q^i \frac{K}{\Mpl^2\,H} - \frac{1}{3H} \dot q^k V_{;k}^i \right|
 \ll \left|3 H \dot{q}^i\right|
  ~.
  \label{consist1}
  \end{equation}
 Here we reapplied the slow roll equations (\ref{slowroll1},\ref{sloweom}) to bring $d^2q^i/dt^2$ to the form (\ref{consist1}).
Dividing both sides by $(3H\dot{q}^i)$, with no sum over $i$, gives
\be
 \left|\frac{K}{V} - \frac{\Mpl^2}{3 V} {\dot q^k V_{;k}^i\over \dot q^i}\right| \ll 1 ~.
\label{consist2}
\ee
In slow-roll, the first term is already small by the smallness of $\hat\epsilon$ (\ref{slowroll1}).     To maintain slow roll, the second term in (\ref{consist2}) must also be small for each field, giving as many conditions as there are fields:
\begin{equation}
\frac{\Mpl^2}{3 V} {\dot q^k V_{;k}^i\over \dot q^i} \ll 1 \quad\forall\,i
\label{slowroll2a}
\end{equation}
The $n$ conditions in \eqref{slowroll2a} give $n-1$ relations between the $\dot q^i$, such that there is only one independent velocity.
It is frequently helpful to have a single parameter to characterize these conditions, particularly when initial conditions are selected so that most of the fields are heavy and most of the kinetic energy is carried by a single field.   A useful parameter is constructed by multiplying the numerator and denominator  of Eq. (\ref{slowroll2a}) by $\dot{q}^i$ and summing over $i$, independently in the numerator and denominator.  This gives a second slow roll parameter
\begin{equation}
\hat \eta \equiv \frac{\Mpl^2}{6 H^2 K} \dot q^i \dot q^j V_{;ij} \ll 1\, .
\label{slowroll2}
\end{equation}
where the constraint on $\hat\eta$ is related to the one remaining constraint in \eqref{slowroll2a}.
The parameter $\hat\eta$ generalizes the conventional single field second slow-roll parameter to the multi-field setting.  The conditions on $\hat\epsilon$ and $\hat\eta$ in (\ref{slowroll1},\ref{slowroll2}) are the first and second slow roll consistency conditions and can be converted using the slow roll equation of motion (\ref{sloweom}) into conditions on the gradient of the potential. Note that in addition to the two slow roll conditions on  $\hat\epsilon$ and $\hat\eta$ we have $n-1$ additional conditions from (\ref{slowroll2a})~\footnote{Requiring that third and higher derivatives of the equation of motion also be small yields further consistency conditions.}. We will return to these when discussing the large volume scenario in the next section.

The number of e-folds for inflation can be evaluated by the integral

\be
N_e = \int_{t_0}^{t_e} H dt~.
\label{Nefolds1}
\ee
We integrate from the time $t_0$ corresponding to 13 e-foldings from the end of inflation to the end of inflation $t_e$. In the space of fields, $q^i$, this is an integral along a path $C$ determined by the dynamics--let this curve be given by some function $f(q^i)=0$. We can then rewrite the integral above in terms of a line integral,
\be
N_e = \int_C {H\over \left|{dq\over dt}\right|}ds\,,\quad
ds=|dq|=\sqrt{G_{ij}dq^idq^j}
\label{Nefolds2}
\ee
For example, if there are only two scalar fields involved and we pick the integration parameter to be $q^1$,
\be
ds = \sqrt{G_{11}}|dq^1|\sqrt{1+2{G_{12}\over G_{11}}{dq^2\over dq^1}+{G_{22}\over G_{11}}
\left({dq^2\over dq^1}\right)^2} \, .
\label{Nefolds3}
\ee 
We will now apply this analysis to the Large Volume Scenario with two K\"{a}hler moduli.

\subsection{Slow roll in the Large Volume Scenario}

This analysis applied to the Large Volume Scenario with two K\"ahler moduli is most transparent in terms of the fields (\ref{diagonalfieldbasis}) which lead to the ``diagonal'' kinetic term (\ref{diagonalkinetic}).    The metric on the field space is then 
\begin{equation}
 G_{11} = 1\,,\quad G_{22} = e^{\sqrt{6}\,q^1/\Mpl}={4\over 3} {\beta_2\over \left(q^2\right)^2}\,\Mpl^2~.
\end{equation}
In this approximation, the inverse components of the field space metric are 
\begin{equation}
G^{11} = 1\,,\quad G^{22} = e^{-\sqrt{6}\,q^1/\Mpl}={3\over 4} {\left(q^2\right)^2\over \beta_2}\,\Mpl^{-2}~.
\end{equation}
Then the two slow roll conditions in (\ref{slowroll2a}) are
\begin{eqnarray}
  \left|V_{;11} + \frac{\dot{q}^2}{\dot{q}^1}\,V_{;12}\right| & \ll & \frac{V}{\Mpl^2}\,.
  \label{acc1}\\
  \left|G^{22}\,V_{;22} + \frac{\dot{q}^1}{\dot{q}^2}\, G^{22}\,V_{;21}\right| & \ll & \frac{V}{\Mpl^2}\,, 
  \label{acc2}
\end{eqnarray}
While there are many ways to satisfy these inequalities, inspection of the potential near the large volume minimum indicates the presence of a valley aligned roughly along the $q^1$ direction while walls in the $q^2$ direction are steep.   This indicates that, given an initial condition in this valley, $q^2$ will be heavy and most of the kinetic energy of rolling will be carried by $q^1$.   Anticipating such an initial condition we solve (\ref{acc2}) by setting it identically to zero.  This relates the velocities in the two fields
\be
{\dot q^2\over \dot q^1}\approx - {V_{;12}\over V_{;22}}~.
\label{slow-velocity}
\ee
(Thus if  the potential is steeply curved in the $2$ direction -- i.e. large $V_{;22}$ -- $\dot{q}^2$ will tend to be small relative to $\dot{q}^1$. As we will see $q^2$ may still give a non-trivial contribution to the observables.)  Equivalently, applying the slow-roll equation of motion \eqref{sloweom},
\be
{V_2\over V_1}\approx -{V_{;12} \over G^{22} V_{;22}}~.
\label{slow-gradient}
\ee

We can now explicitly compute the slow-roll parameters. By using (\ref{slow-velocity}) and both of the slow-roll conditions (\ref{slowroll1}) and (\ref{sloweom}), we get\footnote{Notice that $G^{22}\,V_{;22}\propto \beta_2\ll 1$, using the expressions in appendix A. This is so because the proportionality factor is a function of $(z_i,\,y_0)$ and is of order one once we fit the observables in the next section. Thus, the ratio between the kinetic energy densities, $G_{22}(\dot{q}^2)^2/(\dot{q}^1)^2$, is indeed subdominant.}
\be
\hat\epsilon\approx{\Mpl^2\over 2}\left({V_1\over V}\right)^2\left(1+O(\beta_i)\right)
\label{firstslowroll12}
\ee
Similarly,
by using  (\ref{slow-velocity}) and the first slow-roll condition (\ref{slowroll1}) one finds
\be
\frac{\hat\eta}{\Mpl^2}\approx {V_{11} V_{22} - \left(V_{12}\right)^2\over V \, V_{22}}+ O(\beta_i)
= {\det(V_{ij}) \over V \, V_{22}} + O(\beta_i)
~.
\label{etahat12}
\ee
Here we have used the results for the covariant derivatives of the potential, $V_{;ij}$, calculated in appendix A, and then only kept the zeroth order terms in $\beta_i$.    In this approximation the covariant derivatives can be replaced by simple partial derivatives.  In addition, for both $\hat\epsilon$ and $\hat\eta$, terms of $O(\beta_i)$ have been dropped  in our expansion of the kinetic energy because they make subleading contributions,
\be
K= {1\over 2} \left(\dot q^1\right)^2\left(1+O(\beta_i)\right)~.
\label{kineticexpansion}
\ee
In fact, by dividing \eqref{acc1} by $V/\Mpl$ and using the relationship between the two velocities (\ref{slow-velocity}), we can rewrite the second slow-roll condition (\ref{acc1}), we find
\be
  \left|\frac{\Mpl^2\,V_{;11}}{V}\left(1 - {\left(V_{;12}\right)^2\over V_{;11}\,V_{;22}} \right)\right| \
    \approx \  |\hat{\eta}|  \ \ll \  1\,.
\ee

In order to evaluate the number of e-folds in this scenario using (\ref{Nefolds2}) we need to know the path that the dynamics picks for us. To a good approximation, during slow roll, we have that $f(q^i) \equiv V_2\approx 0$. Thus, using the implicit function theorem, we have that
\be
{dq^2\over dq^1} = -{\nabla_1 f\over \nabla_2 f}=
-{V_{;12}\over V_{;22}}~.
\ee
To evaluate $ds$, we use that the metric is diagonal and that $\beta_2<<1$ and hence obtain that
\be
ds =\sqrt{1+G_{22}\left({V_{;12}\over V_{;22}}\right)^2} \approx dq^1~.
\ee
Similarly, we approximate 
\be
\left|{dq\over dt}\right| = \sqrt{G_{ij} 
\, \dot q^i \, \dot q^j} = \sqrt{2 \, H^2 \, \Mpl^2\hat\epsilon}
\approx H \Mpl^2 \left|{V_1\over V}\right|~.
\ee
Thus, the number of e-folds is given by
\be
N_e \approx \int {V \over \Mpl^2V_1}  dq^1
\ee

For our present purpose of studying whether the inflationary observables can be sensitive to the topology of extra dimensions, we will consider initial conditions that support slow roll, compute a set of cosmological observables in this dynamical regime and then ask how these observables vary as the microscopic parameters change, assuming that the right number of e-foldings is attained, see section 5.

\section{Observables}

In a multi-field inflationary model, there are two gauge invariant notions of quantum fluctuations. We can always decompose these along the direction of classical motion $(Q_\sigma)$ and its transverse direction $(\delta s)$.  The former ones are related to curvature perturbations (the standard adiabatic perturbations of single field inflationary models), whereas the latter are the source for isocurvature perturbations (or entropy perturbations).  We will focus attention on the power spectrum of adiabatic perturbations ${\cal P}$, the associated spectral index $n_s$, its running $\alpha_s$ and the ratio of tensor to scalar perturbations $r$.  In keeping with convention,  our inflationary observables should be evaluated at field values $x_0,y_0$ corresponding to an epoch about $13$ e-foldings before the end of inflation \cite{liddle-lyth}.

\vspace{0.25in}
\noindent {\bf Adiabatic power spectrum: } The power spectrum of adiabatic perturbations is specified as \cite{lalak2007}
\begin{equation}
 <Q_{\sigma\,k}^\star\,Q_{\sigma\,k^\prime}> = \frac{2\pi^2}{k^3}\,{\cal P}(k)\,\delta(k-k^\prime)
\label{powerspectrum1}
\end{equation}
At lowest order in the slow roll parameter expansion and assuming the amplitude of these perturbations does not evolve quickly after Hubble crossing,
\begin{equation}
{\cal P} \approx \left(\frac{H^2}{2\pi \dot\sigma}\right)^2=\frac{H^4}{2 (2\pi)^2 K}\,,\quad {\rm where}\quad \dot \sigma^2 = G_{ij}\dot q^i\dot q^j=2K~.
\end{equation}
Using the slow-roll equations, it can be rewritten as
\begin{equation}
{\cal P} \approx \frac{2}{3\pi^2}\,\frac{V}{\Mpl^4}\,\frac{1}{16 \, \hat{\epsilon}}\, ,
 \label{power}
\end{equation}
where $V$ and $\hat{\epsilon}$ are evaluated at field values $x_0,y_0$ that occur about $13$ e-foldings prior to the end of inflation. Hereafter, ${\cal P}$ denotes ${\cal P}(k_0)$ where $k_0$ is the scale corresponding to $x_0,y_0$. In the large volume scenario, $\hat\epsilon$ is given by (\ref{firstslowroll12}) and hence to lowest order in $\beta_i$ we have
\be
{\cal P}\approx \frac{1}{12\pi^2}\,\frac{V^3}{\Mpl^6 V_1^2}\,.
\label{slowrollpower}
\ee
The power spectrum depends  on the microscopic parameters, i.e., the $z_i$, $x_0$ and $\hat\gamma$, via the potential and its derivative.  The explicit expression can be found in Appendix B.

\vspace{0.25in}
\noindent {\bf Tensor to scalar ratio: } Following Lyth and Riotto \cite{lythriotto} we can read off the ratio of  the amplitudes of the tensor and scalar power spectra from the form of (\ref{power}) as\footnote{If we take into account the non-trivial kernel that propagates the ratio reported here to the one that is actually measured in the CMB, the proportionality between $r$ and $\hat\epsilon$ is different from 16, but
such considerations will not be of relevance for the sensitivity analysis to be reported in section 5.}:
\begin{equation}
r = 16\,\hat\epsilon\, \approx 8\Mpl^2 \left(\frac{V_1}{V}\right)^2\,.
\label{tensor}
\end{equation}
For later purposes, it will be more convenient to use the square root of the above relation, 
\be
s\sqrt{r} = 2\sqrt{2}\Mpl \frac{V_1}{V}\,.
\label{sqrtr}
\ee
where $s=\text{sign}(V_1)$ gives the sign of the slope of the potential.
Using the explicit form of $V$ in our large volume model, we can compute how $\sqrt{r}$ depends on the microscopic parameters, in terms of the $z_i$. The expression is given in Appendix B.

\vspace{0.25in}
\noindent {\bf Spectral Index: }   The spectral index $n_s$ is defined as,
\begin{equation}
  n_s{-}1 = \frac{d\log{\cal P}}{d\log k}\,. 
\end{equation}
Following the computation discussed in appendix \ref{sec:obser}, assuming that slow roll takes place and expanding to lowest order in the slow roll parameters, the spectral index equals :
\begin{equation}
  n_s{-}1  \approx H^{-1} \frac{\dot {\cal P}}{{\cal P}} = - 6 \hat \epsilon + 2\hat \eta\,,
\label{ns}
\end{equation}
where $\hat{\epsilon}$ and $\hat{\eta}$ are given in (\ref{slowroll1},\ref{slowroll2}).
For the case of the large volume model considered above we use that
 these parameters are approximated as (\ref{firstslowroll12},\ref{etahat12}) to leading order in $\beta_i$:  
 \be
 n_s{-}1 \approx \Mpl^2\left(- 3 \left({V_1\over V}\right)^2 + 2 {V_{11}\over V}\left(1-{(V_{12})^2\over V_{11}V_{22}}\right)\right)~.
 \ee
 Notice that despite  $q^2$ carrying a negligible amount of kinetic energy, as seen in (\ref{kineticexpansion}), $\hat\eta$ is {\it not} equal to the slow roll parameter of an effective single scalar field inflationary model, which would have matched the first term $(V_{;11}/V)$ in (\ref{etahat12}). 
This is essentially because  $q^2$ may not have negligible velocity, despite carrying much less kinetic energy. The explicit expression for $n_s {-} 1$ in terms of the microscopic parameters is given in Appendix B.

\vspace{0.25in}
\noindent {\bf Running of the spectral index: }   The running of the spectral index $\alpha_s$ is defined by
\begin{equation}
  \alpha_s = \frac{d n_s}{d\log k} \approx H^{-1}\,\frac{dn_s}{dt}\,,
\end{equation}
where the last step follows from the considerations discussed in appendix \ref{sec:obser}. 
In this appendix, it is also shown that $\alpha_s$ reduces to
\begin{equation}
  \alpha_s \approx -24 (\hat \epsilon)^2+ 16\hat \epsilon\,  \hat \eta - 2 \hat \psi + 4\left((\hat \eta)^2-\widehat{(\eta^2)}\right)
  \label{run} 
\end{equation}
where we defined :
\begin{equation}
\widehat{(\eta^2)} = \frac{1}{18 H^4 K}\, \dot q^j \dot q^k V_{;ik} V_{;j}^i\,,\quad
\hat \psi = -\frac{1}{6 H^3 K}\, \dot q^i\dot q^j\dot q^k \nabla_k\left(V_{;ij}\right)~.
\end{equation}
The appearance of the term $4\left((\hat \eta)^2-\widehat{(\eta^2)}\right)$ is in agreement with a similar computation in \cite{lythriotto}.  A careful examination and evaluation of this last term, reported in appendix \ref{sec:obser}, tells us that for our models, it cancels :
\begin{equation}
  4\left((\hat \eta)^2-\widehat{(\eta^2)}\right) \approx \eO(\beta_0)\,.
\end{equation}
Thus, in our models, $\alpha_s$ is
\begin{equation}
  \alpha_s \approx -24 (\hat \epsilon)^2+ 16\hat \epsilon\,  \hat \eta - 2 \hat \psi\,,
  \label{runningexp}
\end{equation}
where the new slow roll parameter $\hat\psi$, expanded to lowest order in $\beta_i$, is equal to:
\begin{equation}
\hat\psi \approx \Mpl^4\,\frac{V_1 V_{111}}{V^2}\,\left(1-3{V_{112}\over V_{111}} {V_{12}\over V_{22}} +
3 {V_{122}\over V_{111}} \left({V_{12}\over V_{22}}\right)^2 + {V_{222}\over V_{111}} \left({V_{12}\over V_{22}}\right)^3\right)\,.
\end{equation}
As for $n_s$, $\alpha_s$ is {\it not equal} to the naive expression obtained by assuming a single effective direction of rolling, as it also involves the different third order partial derivatives of the potential. The explicit expression for $\alpha_s=\alpha_s(z_i,\,y_0)$ is given in the appendix.

\subsection{Parameters vs. Observables and predicting the tensor-to-scalar ratio}

Above we have derived expressions for four CMB observables $\{{\cal P}, r, n_s, \alpha_s\}$ in terms of derivatives of the potential.   Appendix A evaluates these derivatives in terms of the effective parameters $\{z_i, \hat{\gamma}, x_0,y_0\}$.  All the microscopic parameters, including the topological parameters of the extra dimensions, are encapsulated in the $z_i$, while $x_0,y_0$ are the field values at the epoch of horizon exit for  wavenumber at which the observables are evaluated,
ie 13 e-folds before the end of inflation.
In Appendix B, these computations are assembled into explicit expressions for the observables as a function of the parameters.  Here we invert these relations to ask how the microscopic parameters are determined by measurement of the observables.

First of all, realizing slow roll inflation requires fine tunings that already impose relations between the parameters.  Specifically, we required in (\ref{slow-gradient},\ref{firstslowroll12})  $q^2 V_2 \ll V_1\ll V$.  Thus, in order to achieve slow roll, to lowest order in $\beta_0$ and the slow-roll parameter $\hat\epsilon$, we fine tuned $q^2 V_2\approx 0$, from which it follows that  one of our parameters is no longer independent.   Using the explicit form of $V_2$ given in Appendix A, we get
\begin{equation}
  z_2 = \frac{y_0-1}{4y_0-1}\,z_1 + \eO(\beta_0\,\sqrt{r})\,.
\label{constraint1}  
\end{equation}
Thus it turns out that the observables $\{{\cal P},n_s,\,r,\,\alpha_s\}$ can be expressed entirely in terms of two of our parameters $\{z_1,\,z_3\}$, the initial condition $y_0$ and the sign of the slope of the potential.
 Inverting these relations, and using (\ref{constraint1}) we can express the $z_i$ as
\begin{equation}
z_i (r, n_s{-}1)  \approx z_i^0(y_0) + h_i(y_0) \,  (n_s {-} 1)
+ g_i(y_0) \, s \, \sqrt{r} + {\cal O}(r, (n_s{-1})^2)
\end{equation}
to leading order in the observables.   Here $z_i^0$, $h_i$ and $g_i$ are functions only of the initial condition $y_0$ and $s$ is a sign that depends on whether the inflaton is rolling towards larger or smaller volumes in the extra dimensions.   The complete expressions for $z_i$ are given in Appendix C, with $z_i^0$ given in  Eqs. (\ref{z1limit}-\ref{z3limit}) . To simplify, we can approximate the functional dependence on $y_0$ by using the fact that for the validity of the effective potential we must have $y_0 \gg 1$ since $y_0$ measures the volume, in string units, of a cycle in the extra dimensions.    Indeed, near the minimum of the potential in the Large Volume Scenario, $y_0$ is typically $\eO(10)$ \cite{LVS}.   This gives
\begin{eqnarray}
z_1(r, n_s{-}1) - z_1^0(y_0) 
\ \approx \
z_1(r, n_s{-}1) - {4 \over 9} \, y_0 
\ & \approx & \
 {2 \over 243} \, y_0 \, (n_s {-} 1) 
 -
{1 \over 27 \sqrt{3}} \, y_0 \,  s \, \sqrt{r}
\label{z1leading}
\\  
z_2(r, n_s{-}1) - z_2^0(y_0) 
\ \approx \
z_2(r, n_s{-}1) - {1 \over 9} y_0
\ & \approx & \ 
 {1 \over 486} \, y_0 \, (n_s {-} 1)
 -
{1 \over 108 \sqrt{3}} \, y_0 \,  s \, \sqrt{r}
 \label{z2leading}
\\
z_3(r,n_s{-}1) -  z_3^0(y_0) 
\ \approx \
z_3(r,n_s{-}1) -  {4 \over 27} y_0
\ & \approx & \ 
 {2 \over 729} \, y_0 \, (n_s {-} 1)
 -
{10 \over 243 \sqrt{3}} \, y_0 \,  s \, \sqrt{r}
 \label{z3leading}
\end{eqnarray}
where in the middle we have given the leading term in $z_i^0$ from Eqs. (\ref{z1limit}-\ref{z3limit}) and on the right hand side we have displayed the leading terms in $h_i$ and $g_i$.      Note that the $z_i$ are all of $\eO(1)$, and hence the smallness of $\hat\epsilon \approx (\Mpl^2/2) (V_1/V)^2$, as required by the slow roll conditions, occurs because of cancellations between the different terms in $V_1$. Similarly, for $\hat\eta$ we have that $\det\left(V_{ij}\right)$ is relatively small, or in other words that the field that rolls is light. 

Having solved for the $z_i$ in terms of $\{n_s,r,y_0\}$,  we can insert these quantities back into the expressions for the observables to find possible relations between them.  Specifically, starting with (\ref{runningexp}) for the running of the spectral index, we can subtitute in terms of derivatives of the potential, and then use the expressions in Appendices A and C to express $\alpha_s$, first in terms of the $z_i$, and then in terms of $\{n_s, r, y_0\}$.  To leading order 
\be
\alpha_s  \approx   23 \, s \, \sqrt{r}   
\label{leadingrunning}
\ee
where we used the current WMAP value $n_s \sim 0.95$ \cite{WMAP3}, and the fact that $y_0 \gg 1$
in making estimates of the numerical coefficients.  The sign $s$ of the running of the spectral index arises from the sign of the slope of the potential and hence depends on whether the volume of the extra dimensions is increasing or decreasing due to the rolling of the inflaton.   
Hence, requiring that $|\alpha_s| < 0.1$ as per current bounds, predicts a scalar to tensor ratio $r$ that is far below the  $10^{-3}-10^{-4}$ band of nearby future technological accuracy \cite{VPJ}.   Thus, if the effects of primordial gravitational waves are measured in the near future, this model is ruled out.

The definitions of the effective parameters $\{z_i\}$ in terms of the microscopic ones given in \eqref{z1}-\eqref{z3} combined with  the leading order expressions in (\ref{z1leading}-\ref{z3leading}) allow us to derive the order of magnitude relations
\begin{equation}
  y_0\frac{A\,e^{-y_0}}{|W_0|} \sim \frac{\xi\,x_0}{2}  \sim \frac{3}{4}\,\beta_0
   \sim \frac{8\pi}{5}\frac{\hat{\gamma}\,y_0}{g_s^4\,|W_0|^2} \, .
   \label{microrelations}
\end{equation}
These are consistent with our initial analytical approximations \eqref{eq:approx}.

While the observables $\{n_s, r, \alpha_s\}$ constrain the parameters $z_i$ and the initial condition $y_0$ as above, the parameter $\hat{\gamma}$ and the initial condition $x_0$ are constrained only by the power spectrum ${\cal P}$.  Using the expression (\ref{powerz}) in Appendix B for ${\cal P}$, and recalling that $r = 16 \,\hat{\epsilon}$ (\ref{tensor}), we find that
\begin{equation}
{\cal P} \approx {2\over 3 \pi^2} {\hat\gamma x_0^2\over r} \left(1+3z_3+4 z_2-2 z_1\right)\approx 7.5\, 10^{-3}
{\hat\gamma x_0^2\over r} 
\end{equation}
where in the last approximation we have used the  explicit expressions for the $z_i$ from Appendix B, considered the leading small $r$, large $y_0$ limit, and evaluated the spectral index at the current WMAP value $n_s\sim 0.95$.
Using (\ref{microrelations}) to relate $\hat{\gamma}$  with $x_0$  (up to a dependence on $y_0$, $\xi$ and $g_s^4\,|W_0|^2$), we note that the volume of the extra dimensions ($x_0^{-1}$) is observationally determined by the amplitude of the power spectrum, ${\cal P}$, and the tensor to scalar ratio, $r$.

\section{Observable dependence on microscopic parameters}

Our main purpose in this paper is to assess whether cosmological observables might be able to constrain the topology of extra dimensions.  Let us first enumerate the parameters in the simple two-field model that was developed in detail above.     The microscopic parameters of this simplified theory are the Euler number $\chi$, the intersection numbers $\lambda$ and $\mu$ and the four superpotential coefficients $\{a_i,\, A_i\}$.  While we have been treating these quantities as independent variables, in fact the choice of a particular Calabi-Yau manifold  with two K\"ahler moduli as a description of the extra dimensions determines all the parameters $P = \{ \chi, \lambda, \mu
\}$, i.e. these parameters are not independently tunable; rather, they jointly vary in a discrete and sparse way between manifolds.   In addition, the model builder has a choice of ``fluxes'' that may be included in (\ref{eq:potentials}), giving rise to the constant $W_0$ in the superpotential (\ref{eq:potentials2}).   Effectively, this constant is  continuously tunable.    The parameters $\{a_i, A_i\}$ are determined by the nonperturbative dynamical process giving rise to these terms in the superpotential, and are affected both by the discrete choice of topology of the manifold and by the effectively continuous choice of fluxes.    While $a_i$ can only take discrete values,  $A_i$ can effectively vary continuously over some range in a manner that is correlated with the value of $W_0$.   The precise range is not understood, since the $A_i$ have not been explicitly computed in any significant examples.  Finally, our simple model has two continuously tunable initial conditions $x_0$ and $y_0$.  

 Any model of slow roll inflation will have to satisfy the slow roll conditions (\ref{epsilonhat}, \ref{slowroll2a}), and we could use $x_0, y_0$ to achieve this.   At  the third order in slow roll, i.e. requiring that $d^3 q^i / dt^3$ be small enough, there are two more slow roll conditions.  (We have not specified these explicitly in our analysis, but they are related to the smallness of the running of spectral index $\alpha_s$ and of  $\hat\psi$ in (\ref{runningexp}).)  Possibly these two conditions can always be satisfied by tuning the $A_i$, although it is not certain that the range over which $A_i$ can be tuned would be wide enough.  Matching one additional observable could be achieved by tuning $W_0$.     But matching all other observables, and requirements like achieving the right number of e-foldings,   have to be accomplished by a choice of topology of the extra dimensions and the $a_i$.  Since the remaining parameters, $P$ and $a_i$, vary discretely and sparsely, this will generally be difficult, and in our simple two-field example would require conspiracies between the parameters.  Because of this, it is clear that cosmological observations will be highly constraining of the topology of the simple two-field example.

Generically, a string theoretic model of inflation will have many scalar fields parameterizing the geometry of the extra dimensions.  Indeed, when a model builder changes the topology of the extra dimensions he or she will generally change the number of low energy scalar fields.   Increasing the number of fields will increase the number of initial conditions that can be tuned.  Hence we should ask whether the above conclusion regarding sensitivity of the model to the topology of the extra dimensions will apply generally.  A general Calabi-Yau geometry with $N$ low-energy scalar fields will have parameters $\mathbb{P} =\{\chi, \lambda_{ijk}, a_i, A_i\}, \ i,j,k = 1, 2 \dots N$ where $\chi$ is the Euler number, $\lambda_{ijk}$ are the intersection numbers of the geometry and $\{a_i,\,A_i\}$ are the dynamically determined parameters appearing in the superpotential.   As in the two-field case, $\chi$, $\lambda_{ijk}$, and $a_i$ will be discrete and sparse, although $a_i$ will depend on the choice of ``fluxes" in the extra dimensions.   As in the two-field case, these ``fluxes'' will also give rise to a continuously tunable constant $W_0$ and will cause the $A_i$ to be continuously tunable over some (as yet unknown) range.  Finally, there are $N$ initial conditions $x_0,y^i_0\,,i=2,\ldots,N$.  However, following (\ref{epsilonhat}) and (\ref{slowroll2a}) there will now be $N$ 
second order slow-roll conditions that will fix  the initial conditions.   We should expect that the third order slow-roll conditions will constrain the continuously tunable $A_i$.   Then $W_0$ can be fixed by requiring that one of the observables is matched.    All remaining observables, and requirements like achieving the right number of e-foldings, the reheating temperature etc. must then be achieved by the choice of topology and $a_i$.   Because of the  discreteness and sparseness of the remaining parameters, this will generally be difficult to achieve.   This implies that cosmological observables will constrain the topology of the extra dimensions in these scenarios.

In order to get a concrete grasp on these considerations we will examine the details in the simplified two-field example that we have been developing throughout this paper.

\subsection{Variation of observables in a two-field model}
To investigate how observational constraints on cosmological observables can be interpreted as constraints on the microscopic parameters, we can equivalently
study the sensitivity of the cosmological observables towards variations in microscopic parameters and the initial conditions.  Our approach will be to assume that slow-roll inflation is realized in some model, and that the observables have been evaluated at the correct epoch about 13 e-foldings before the end inflation.  Then we will vary the microscopic parameters and ask how sensitive the observables are to the variations in a linearized approximation.  This is similar to the Fisher matrix analysis sometimes used in cosmology.   Of course, the variations in parameters must preserve the slow roll regime, and also the  set of analytical approximations that allowed us to simplify the effective 4d potential in the Large Volume Scenario.    This provides constraints among the different variations.   We are most interested in the variations in the parameters that arise from changing the topology of the extra dimensions, which lead here to changes  in the Euler number and the intersection numbers.   

In order to preserve the slow roll conditions we have to make sure that
\bea
\delta\left(\hat\epsilon\right) &\approx& 3\,\delta\left({K\over V}\right) <<1\\
\delta\left(\frac{d^2q^i}{dt^2}{1\over 3H\dot{q}^i} \right) &\approx& \delta\left({K\over V} \right) - 
\delta\left({M_P^2\over 3 V}{\dot q^kV_{;k}^i \over \dot q^i} \right) <<1\quad\forall\,i~.
\label{variationslowroll}
\eea
The former constraint amounts, in the large volume scenario, to
\begin{equation}
  \delta\left(\frac{K}{V}\right) \sim 2\frac{K}{V}\,\left(\frac{\delta V_1}{V_1} -  \frac{\delta V}{V}\right)\ll 1\,.
\end{equation}
where we suppressed subleading terms of $\eO(\beta_i)$. This is achieved simply by requiring :
\begin{equation}
  \left|\frac{\delta V_1}{V_1} - \frac{\delta V}{V}\right| \leq 1\,,
\end{equation}
due to the overall suppression caused by $K/V$.
Since the tensor to scalar ratio $r=16\,\hat\epsilon$, requiring that $\delta\left(\hat\epsilon\right)\ll 1$ amounts to make the variations in $r$ small as well.

As discussed earlier, the second set of slow roll conditions, can in the case of our large volume scenario be rewritten in terms of the constraint $|\hat\eta|<<1$ and one further constraint coming from the heavy, second field. Thus, the two constraints in \eqref{variationslowroll} give the following conditions
\bea
\delta\left(\hat\eta\right) &\approx& \hat\eta\left(-{\delta V\over V} - {\delta V_{22}\over V_{22}} +{\delta\left(V_{11} V_{22} - (V_{12})^2\right)\over V_{11} V_{22} - (V_{12})^2}\right) \ll 1 \label{variationslowroll2a} \\
  \delta\left({M_P^2\over 3 V}{\dot q^kV_{;k}^2 \over \dot q^2} \right) &\approx& \delta\left[\frac{\Mpl^2}{3}\,\frac{G^{22}\,V_{22}}{V}\,\left(1 + \frac{\dot{q}^1}{\dot{q}^2}\,\frac{V_{12}}{V_{22}}\right)\right]\ll 1\,,
 \label{variationslowroll2b} 
\eea
where we suppressed subleading terms of $O(\beta_i)$. Because $|\hat\eta|<<1$, we satisfy \eqref{variationslowroll2a} by requiring
\be
\left|-{\delta V\over V} -{\delta V_{22}\over V_{22}} +{\delta\left(V_{11} V_{22} - (V_{12})^2\right)\over V_{11} V_{22} - (V_{12})^2}\right| \leq 1~.
\ee
Since the spectral index, $n_s{-}1$, depends on $\hat\eta$ (and $\hat\epsilon$), ensuring that 
$\hat\eta<<1$ will guarantee that $n_s{-}1$ is not allowed to vary too much, as is usual in slow roll
inflation.

Finally, we turn to \eqref{variationslowroll2b}.
The variation can be split into two natural contributions. The first one, coming from $\delta \left(G^{22}\,V_{22}/V\right)$, will vanish because it is multiplied by the second factor which vanishes, by assumption,
due to slow roll. We are thus left with the second term :
\begin{equation}
  \delta\left({M_P^2\over 3 V}{\dot q^kV_{;k}^2 \over \dot q^2} \right) = \frac{\Mpl^2}{3}\,\frac{G^{22}\,V_{22}}{V}\,\delta\left(\frac{V_1}{G^{22}\,V_2}\,\frac{V_{12}}{V_{22}}\right)\,,
\end{equation}
where we used the slow roll equations of motion for $\dot{q}^i$. Notice that this variation equals :
\begin{equation}
  \delta\left({M_P^2\over 3 V}{\dot q^kV_{;k}^2 \over \dot q^2} \right) = -\frac{\Mpl^2}{3}\,\frac{G^{22}\,V_{22}}{V}\,\left(\frac{\delta V_1}{V_1} + \frac{\delta V_{12}}{V_{12}} - \frac{\delta G^{22}}{G^{22}}-\frac{\delta V_2}{V_2}- \frac{\delta V_{22}}{V_{22}}\right)\,,
  \label{variationslowroll2c}
\end{equation}
where we used \eqref{slow-velocity}. Furthermore, our earlier analysis shows that the prefactor
equals
\begin{equation}
 \Mpl^2 \frac{G^{22}\,V_{22}}{V} \approx \frac{64(17+n_s)\, y_0^3}{\beta_0} \gg 1\,.
\end{equation}
We are thus left to conclude that the variations in the second bracket of the RHS of \eqref{variationslowroll2c} have to be very small.
Actually, applying the same analytical philosophy we used for the determination of the slow roll condition \eqref{slow-velocity}, we could just impose, for analytical simplicity, that the variation inside the bracket vanishes. This is equivalent to the geometrical requirement that all variations should preserve
the defining slow roll constraint \eqref{slow-velocity}.  Thus, we can assume
\begin{equation}
  \frac{\delta V_1}{V_1} + \frac{\delta V_{12}}{V_{12}} = \frac{\delta G^{22}}{G^{22}}+\frac{\delta V_2}{V_2}+\frac{\delta V_{22}}{V_{22}}\,.
  \label{var1}
\end{equation}
Note that when satisfying this relation, we have to use that $V_2 = \eO(\beta_2)$ from the slow roll constraint \eqref{slow-gradient}. Thus, ${\delta V_2} = \eO(\beta_2)$ in order to satisfy \eqref{var1}, since the leading order terms in \eqref{var1} are of $\eO(1)$.

Using the expressions for $G_{22}$ \eqref{kineticdiagonal} and the derivatives of the potential reported in appendix A, 
we see that the variations of the slow roll conditions 
only depend on the set of parameters $\{z_i,\,y_0,\,\beta_0,\,\hat\gamma\,x_0^2,\,\mu\,x_0\}$ 
\footnote{The dependence on $\mu\,x_0$ follows from writing $G_{22}=4\beta_2/(3 q^2)$ where $q^2 = M_p \sqrt{\beta_2} \,\mu x\, 2/\sqrt{3}$
and evaluating at the start of inflation.}.
These are not fundamental parameters, but they will clearly encode any variation of the fundamental parameters of the theory. Note that all the dependence on
$\hat\gamma\,x_0^2$ is universal in the variation of the slow-roll constraint equation \eqref{var1}: all the potential derivatives appearing in \eqref{var1}, when viewed
as functions of $\hat\gamma\,x_0^2$ are homogeneous of degree one. 
The same applies to variations of $\mu\,x_0$. This is because the variation through $G^{22}$ exactly cancels the contributions from $\delta q^2$ in the different potential derivatives $\{V_2\,,V_{12},\,V_{22}\}$. Thus, we conclude that \eqref{var1} is independent of the variations with respect to $\hat\gamma\,x_0^2$ and $\mu\,x_0$, and hence there are only five main variations to consider in the parameters $\{z_i,\,y_0,\,\beta_0\}$.   We know turn to the variations of the observables.

\vspace{0.25in}
\noindent
{\bf Adiabatic power spectrum: }
The power spectrum varies as :
\begin{equation}
  \frac{\delta\eP}{\eP} = 3\,\frac{\delta V}{V} - 2\frac{\delta V_1}{V_1}\,.
  \label{apsvariation}
\end{equation}
Thus, essentially the same functions controlling $K/V$ are also responsible for the variations of $\eP$. Because the adiabatic power spectrum has been measured accurately we have a tight constraint on
$\delta{\cal P} \approx 0.05\,{\cal P}$ at the one sigma level and using data only from WMAP \cite{WMAP3, VPJ}, and assuming absence of isocurvature power spectrum, no tensor to scalar ratio and no running. To lowest order, and for analytical simplicity purposes, if we require that $\delta{\cal P}\approx0$, we obtain a degeneracy direction given by
\be
\frac{\delta V}{V} \approx {2\over 3}\frac{\delta V_1}{V_1}\,.
\label{apsconstraint}
\ee
This provides one further constraint among the variations of the set of parameters 
$\{z_i,\,y_0,\,\beta_0\}$ and $\{\hat\gamma\,x_0^2,\,\mu\,x_0\}$. 

\vspace{0.25in}
\noindent 
{\bf Tensor to scalar ratio: }
Since $r=16\,\hat\epsilon$ we already know the variation of the tensor to scalar ratio,
\be
{\delta r\over r} = 2\,\frac{\delta V_1}{V_1} - 2\frac{\delta V}{V}\,.
\label{rvariation}
\ee
Using the constraint from the variation of $\eP$ \eqref{apsconstraint} then gives
\be
{\delta r\over r} \approx {2\over 3}\frac{\delta V_1}{V_1}\,.
\label{rconstraint}
\ee
Currently there is only an upper bound on the value of $r$, and hence $\delta r$ is only constrained by the fact that we want to stay in a slow-roll regime.
If the effects of gravitational waves are observed, 
this will then provide yet another constraint among the variations of the set of parameters 
$\{z_i,\,y_0,\,\beta_0,\,\hat\gamma\,x_0^2,\,\mu\,x_0\}$.

\vspace{0.25in}
\noindent {\bf Spectral Index: } 
Recent observations are consistent with the generic inflationary prediction of a non-trivial spectral index, ie $n_s{-}1\neq 0$. Thus the variation of  $n_s{-}1$ is constrained
\be
\delta\left(n_s{-}1\right) \approx -6\delta\hat\epsilon + 2\delta\hat\eta\ll 1~.
\label{nsvariation}
\ee
Future missions (e.g. Planck satellite \cite{planck}) 
are expected to significantly decrease the uncertainty in the deviation from a
scale invariant spectrum to be $\delta\left(n_s-1\right) \approx 0.008$. 
This would allow us to establish a relationship between $\delta\left(n_s{-}1\right)$ and $\delta\hat\eta$, up to  $\delta\hat\epsilon=\hat \epsilon\, \delta(r)/r$. Combined with the constraint from the variation of the adiabatic power spectrum, we would get (assuming that $\delta(n_s{-}1)\approx 0$)
\be
-6\delta\hat\epsilon +2\delta\hat\eta\approx 0\quad\Rightarrow\quad
-4 \hat\epsilon {\delta V_1\over V_1} + 2\hat\eta\left(-{2\over 3} {\delta V_1\over V_1} -{\delta V_{22}\over V_{22}} +{\delta\left(V_{11} V_{22} - (V_{12})^2\right)\over V_{11} V_{22} - (V_{12})^2}\right)\approx 0~.
\label{nsconstraint}
\ee
This will then provide yet another degeneracy direction for the variations of the set of parameters 
$\{z_i,\,y_0,\,\beta_0,\,\hat\gamma\,x_0^2,\,\mu\,x_0\}$.

\vspace{0.25in}
\noindent {\bf Running of the spectral index: }
The final observable is $\alpha_s$, which is second order in the slow roll parameters $\hat\epsilon$ and $\hat\eta$, and also depends on the purely second-order slow-roll parameter $\hat\psi$. We have that
\be
\delta\alpha_s = - 48\, \hat\epsilon^2 \,{\delta\hat\epsilon\over \hat\epsilon}  + 16\, \hat\epsilon\,\hat\eta\left({\delta \hat\epsilon\over \hat\epsilon} + {\delta \hat\eta\over \hat\eta}\right)
-2 \,\delta\hat\psi~.
\label{asvariation}
\ee
Although the experimental bound on $\alpha_s$ is less severe, WMAP alone suggests $\delta \alpha_s\approx 0.03$, though it can grow up to $\delta\alpha_s \approx 0.043$ for non-zero $r$. At any rate, this
provides yet another constraint among the variations of the set of parameters 
$\{z_i,\,y_0,\,\beta_0,\,\hat\gamma\,x_0^2,\,\mu\,x_0\}$.

We will now combine the constraints discussed above to  consider what happens when the Euler number is varied. A variation of the Euler number implies that the topology of the extra dimensions has changed, which means that  $\chi$ will not vary alone.  In general the intersection numbers will change, and since the complex structure of the manifold will be different, the fluxes $G_3$ in  (\ref{eq:potentials}) and resulting constant $W_0$ in (\ref{eq:potentials2})  will change also.   Similarly the constants $A_i$ and $a_i$ appearing in the exponential terms in the superpotential (\ref{eq:potentials}) will change to values determined by the structure of the new manifold.    The details of these joint variations are model dependent, so for simplicity we will simply assess the sensitivity of the observables to joint variations in the Euler number $\chi$, the tunable parameter $W_0$, and the initial conditions $x_0,y_0$.

\subsection{Euler number variations in a two-field model}

First, consider variations in the Euler number, keeping all other parameters and initial conditions fixed. Since the Euler number only appears in $z_3$, we will focus on $\delta z_3$ variations. These satisfy, using the relationship between $z_3$ and the microscopic parameters \eqref{z3} derived in section II.A. for the two-parameter model :
\begin{equation}
  \delta z_3 = z_3\,\frac{\delta\xi}{\xi}\,,
\end{equation}
due to the linear dependence of $z_3$ on $\xi$.

We can now check how a small variation of $\xi$ alone will be inconsistent with preservation of the slow roll regime. The consistency condition~\eqref{var1} simplifies considerably in this case to $\delta V_1/V_1 \approx 0$,  since only $\delta V_1$ is non-vanishing. Here
\be
{\delta V_1\over V_1} = -{9 z_3\over -2 + 4y_0(z_1-4 z_2)- 9 z_3} 
{\delta z_3\over z_3}\approx -{4 (17+ n_s) y_0\over\sqrt{3}\sqrt{r}} \frac{\delta\xi}{\xi}~.
\ee
and hence since the minimal change in the Euler number is $|\delta\chi|=2$ and $\delta \xi/\xi \geq 10^{-3}$   (where we used that $\xi \, < \,\eO(10^3)$ for the known CYs), it is clear that we are no longer in the slow-roll regime.

As anticipated, to stay in the slow roll regime requires a fine tuning of the microscopic parameters. The above example shows that even the most delicate fine tuning is not sufficient if {\it only} the Euler number is varied. Let us therefore consider a more general  variation in which the Euler number $\chi$ and the flux contribution to the superpotential $W_0$ are jointly varied.

From the relationship between $z_i$ and the microscopic string parameters we have
\be
{\delta z_1\over z_1} = {\delta |W_0|\over |W_0|}\,,\quad \delta z_2 =0\,,
\quad {\delta z_3\over z_3} = 2  {\delta |W_0|\over |W_0|} + {\delta \xi\over \xi}\,,\quad
\delta G_{ij}=\delta G^{ij}=0~.
\ee
The slow-roll constraint \eqref{var1} then simplifies to
\be
{\delta V_{22}\over V_{22}}-{\delta V_1\over V_1}-{\delta V_{12}\over V_{12}} + {\delta V_2\over V_2}=0~.
\ee
This will lead to a constraint between $\delta|W_0|$ and $\delta \xi$, 
which in the leading small $r<<1$ and large $y_0>>1$ limit becomes
\be
{\delta |W_0|\over |W_0|} \sim -{27 \beta_2\over 8 y_0} {\delta \xi\over \xi}~.
\ee
Note that since $\beta_2<<1$, this amounts to a fine tuning of $|W_0|$ in order for slow-roll to continue to take place after the variation of the Euler number.  Using this constraint we then find that
\be
{\delta\hat\epsilon\over \hat\epsilon} \approx 2 \left({\delta V_1\over V_1} -{\delta V\over V}\right) \sim -{8(17+n_s) y_0\over \sqrt{3} \sqrt{r}} {\delta \xi\over  \xi} \geq \eO(1).
\label{epsilonvariationchi}
\ee
since $r<1$, $\delta \xi/\xi \geq 10^{-3}$ and $y>1$. As long as $\delta\hat\epsilon/\hat\epsilon$ is not much larger than one, slow roll will still take place.

We now turn to the sensitivity of the observables to the above variations. First, 
the variation of the power spectrum \eqref{apsvariation} becomes
\be
{\delta{\cal P}\over {\cal P}} 
\sim {8(17+n_s) y_0\over \sqrt{3} \sqrt{r}} {\delta \xi\over  \xi} \geq \eO(1)~.
\ee
Since $\delta{\cal P}/{\cal P}$ is of the order of $0.05$, variations in $\xi$ are tightly constrained. Similarly,
the variation of the tensor to scalar ratio \eqref{rconstraint} becomes
\be
{\delta r\over r} 
\sim -{8(17+n_s) y_0\over \sqrt{3} \sqrt{r}} {\delta \xi\over  \xi} \geq \eO(1)
\ee
where we have used the result of \eqref{epsilonvariationchi}. Thus, 
we have the somewhat surprising result that the variations of both ${\cal P}$ and $r$ depend very sensitively to changes in the Euler number. In particular, $\delta{\cal P}/{\cal P}$ is outside the current experimental bound when $\delta\xi/\xi \geq 10^{-3}$.
Second, 
in the limit of small $r$ and $y_0>1$ one finds that the variation of the spectral index is
\be
\delta (n_s{-}1) \sim y_0 {(17+n_s)(28-n_s)\over 9}{\delta \xi\over \xi}~.
\ee
Even with $\delta \xi/ \xi \sim 10^{-3}$ and $y_0>1$ this would lead to a change in $n_s{-}1$ that would take us outside the current experimental bound. In fact, the large $\delta (n_s{-}1)$ comes from a large variation of $\hat\eta$. Thus,
 including the variation of the flux contribution to the superpotential, $\delta |W_0|$, in addition to the Euler number variation $\delta \chi$, we are still violating the slow roll condition.

Let us also allow variations of the initial conditions $x_0,y_0$. We find that to leading order in $r$ and for large $y_0>1$
\bea
 {\delta{\cal P}\over {\cal P}}&\sim& {8(17+n_s) y_0\over \sqrt{3} \sqrt{r}} \left(y_0{\delta \xi\over  \xi}+ {(26+n_s)\over (17+n_s)}{\delta x_0\over x_0}-
  {4\over 3} y_0 \delta y_0 \right)
  \label{P-variation-general}\\
   {\delta r\over r}&\sim& - {8(17+n_s) y_0\over \sqrt{3} \sqrt{r}} \left(y_0{\delta \xi\over  \xi}+ {(26+n_s)\over (17+n_s)}{\delta x_0\over x_0}-
  {4\over 3} y_0 \delta y_0 \right)
  \label{r-variation-general}\\
 \delta (n_s{-}1) &\sim&  (17+n_s)\left[{16\over 3} y_0^2\delta y_0 
+ {(28-n_s)\over 9}y_0{\delta \xi\over \xi} + 
 \left({(26+n_s)(13-n_s)\over 6(17+n_s)} \right){\delta x_0\over x_0} \right]\,.
 \label{ns-variation-general}
 \eea
 Thus, requiring $\delta{\cal P}/{\cal P}$ to vanish we also have that $\delta r/r$ has to vanish, which is accomplished by fine tuning the variation of $y_0$ and $x_0$ to satisfy 
 \be
 \delta y_0\sim {3\over 4}\left(  {\delta\xi\over\xi}+{(26+n_s)\over (17+n_s)y_0} {\delta x_0\over x_0}\right)~.
 \ee
 This allows us to simplify the variation of the spectral index,
\be
 \delta (n_s{-}1) \sim 4 y_0 (17+n_s)\left(y_0 {\delta \xi\over \xi}+ {(26+n_s)\over (17+n_s)} {\delta x_0\over x_0}\right)~.
 \ee
since the leading contribution now comes from $\delta y_0/y_0$ and we have assumed that $y_0>>1$. Thus, by fine-tuning the initial conditions $x_0$ and $y_0$ we could remain within an experimental bound on the power spectrum, as well as the bounds on $r$ and $n_s-1$ even after  varying the Euler number.   However, there are additional constraints on inflationary models and observables -- e.g., the running of the spectral index, the number of e-foldings, the requirement that the observables be evaluated at the correct pivot point, the reheating temperature, CMB polarization, and  the isocurvature spectrum.  Given that the tunable parameters $x_0,y_0,W_0$ have been used to maintain slow roll while fixing three observables, {\it all} the remaining constraints have to be achieved by the choice of topology, which we simplified here as amounting to a choice of the Euler number.   This will provide a significant constraint -- indeed, in our simplified two-field setting, given the discrete set of allowed choices of $\chi$, there may be no models at all that meet all the requirements.

The surprising sensitivity of the observables to the topology of the extra dimensions has two sources.  First, the functional form of the potential is highly constrained by its origin in string theory.   Second, topological parameters like the Euler number cannot be fine tuned -- they vary discretely.  We found that such discrete variations can be compensated for in a few observables by fine tuning the effectively continuous parameters like $W_0$ and the initial conditions, but that leaves the discrete choice of topology to account for every other measurable quantity .   In fact, we have been conservative in supposing that  $\delta \chi/\chi$ can be varied at $\eO(10^{-3})$ while holding everything else fixed.  In general, varying the Euler number will entail variations in the intersection numbers also.    Also, since the coefficients $A_i$ and $a_i$ appearing in the exponential contributions to the superpotential (\ref{eq:potentials2}) are dynamically determined, varying the topology changes these quantities too.   Taken together, these parameters form a rather sparse set, at least when evaluated in the most common class of candidate extra dimensional geometries -- the Calabi-Yau manifolds.   Given the degree of fine tuning required to match even present-day measurements, these considerations suggest that cosmological observables could significantly constrain or falsify string theoretic models.

\section{Discussion}

The detailed analyses of this paper focused on K\"ahler inflation in the Large Volume Scenario in a simplified setting involving two fields.  However, our methods and the overall lessons apply more generally:
\begin{enumerate} 
\item The generic inflationary setting in string theory will involve many scalar fields with a field-dependent kinetic term.     The slow-roll parameters and observables of multi-field inflation will generally {\it not} reduce to those of single-field inflation even if most of the kinetic energy is stored in one field.   This is because, particularly in situations with field dependent kinetic terms,  velocities may not be negligible even if the kinetic energy is small. In Section 3 we developed a general methodology for analyzing such settings. This includes the definition of multi-field slow-roll parameters encoding
this important dynamical feature. These are natural extensions of the standard single field inflationary slow roll parameters when there exists an effective rolling direction in field space carrying most
of the kinetic energy. The conditions preserving such dynamical regime were also spelled out and
played an important role in the sensitivity analysis of the cosmological observables.
\item The inflationary observables will be related to the microscopic parameters through effective combinations in which initial conditions, continuous parameters and discrete parameters appear together.  Thus there will necessarily be degeneracies in determining the parameters from the observables.
\item Despite the presence of many parameters, models arising from string theory are highly constrained in their functional form.  Hence, the fine tuning needed to achieve slow roll inflation consistent with observations is difficult to achieve in any specific model.   Also, as a result, the observables can be surprisingly sensitive to the microscopic parameters. 
\item Discrete parameters such as topological numbers of the extra dimensions are particularly interesting because they cannot be finely tuned.   For example, the Calabi-Yau manifolds that typically appear as the extra dimensions of string theory have Euler numbers and intersection numbers that are sparsely spaced integers.  Thus, even coarse constraints from cosmology, limiting the range of these number to within, say, ~500, would be useful in falsifying string theoretic models of inflation, since such constraints have the potential to eliminate most candidate scenarios.   What is more, cosmological observations constrain the shape of the effective potential far from the minimum, thus probing a very different region of theory than the Large Hadron Collider which will provide information about quadratic and cubic terms in the Lagrangian, expanded around its minimum.
\item  In this paper we took as observables the adiabatic power spectrum ${\cal P}(k)$, the tensor-to-scalar ratio $r$, the spectral index $n_s$ and its running $\alpha_s$.    We could have also required that the model predict additional quantities such as the isocurvature power spectrum, the CMB polarization, the current dark energy density, and the reheating temperature.   We did not do this because for simplicity we worked with a two scalar field model with a limited number of parameters, which combined with the  highly constrained functional form for the potential given by string theory, is unlikely to meet all these constraints.  However, the generic model in string theory will contain many more scalar fields, giving more parameters and initial conditions which will then be mutually constrained by the requirement of matching the slow roll observables.
\end{enumerate}
It would be useful to systematically apply the kinds of analysis described here to the many different models of inflation that have been developed in string theory to see whether discrete and topological parameters can generally by constrained by cosmological measurements. 

\acknowledgments
JS would like to thank the Department of Physics at the University of Pennsylvania and at the University of California in Berkeley for their long hospitalities during the completion of this work. PB thanks the organizers of the workshops on ``Generalized Geometry and Fluxes" at DESY and ``String and M Theory Approaches to Particle Physics and Cosmology" at the Galileo Galilei Institute for Theoretical Physics in  Firenze, as well as the University of Tokyo, for their hospitality during various stages of this work. LV is supported in part  by  FP7-PEOPLE-2007-4-3-IRGn202182 and in part by NASA grant ADP03-0000-0092 .  The work of JS was partially supported by the DOE under grant DE-FG02-95ER40893, by the NSF under grant PHY-0331728 and by DOE under the contract number DE-AC02-05CH11231.  VB was supported in part by the DOE under grant DE-FG02-95ER40893, and in part as the Helen and Martin Chooljian member of the Institute for Advanced Study. PB was supported by the NSF grant PHY-0355074 and CAREER grant PHY-0645686.


\appendix
\section{Scalar potential identities}
\label{sec:4d}

In this appendix, we summarise some of the properties of the 4d effective potential responsible for slow roll inflation. This potential is given by :
\begin{equation}
V = \Mp^4 \hat\g x_0^2\Big(\frac{x^2}{ x_0^2}  + 3  z_3\frac{x^3}{x_0^3}
+ 4  z_2 \frac{\sqrt{y}}{\sqrt{y_0}}\,\frac{x}{x_0}\, e^{2(y_0-y)} -
2 \frac{x^2}{x_0^2}\,\frac{y}{y_0}\, e^{y_0-y}  z_1\Big)\,.
\end{equation}

It is the value of this potential and its partial derivatives at the start of inflation $(x_0,\,y_0)$ that will determine most of the observables we will be analysing in this article. 
Although the potential is naturally expressed as a function of $x$ and $y$, measuring the inverse volume of the Calabi-Yau and size of the small 4-cycle, respectively, the inflationary dynamics is computed in terms of the fields $q^{1,2}$ in terms of which the kinetic energy is diagonal.

We first evaluate the (covariant) derivatives to $O(\beta_i)$. To do so we need to compute the connection coefficients $\Gamma^i_{jk}$, which to leading order in $\beta_i$, are given by
\bea
\Gamma^1_{11}&{\approx}& {1\over \Mp}\left(\beta_1 {5\sqrt{3}\over 8\sqrt{2}} {-}\sqrt{6} \beta_2\right),\,\,
\Gamma^1_{12}{\approx} {-}\beta_2 {2\over q^2},\,\,
\Gamma^1_{22}{\approx} {-}M_P \beta_2 {2\sqrt{2}\over \sqrt{3} (q^2)^2}\,,\\
\Gamma^2_{11}&{\approx}& {1\over M_P^2}\left({3q^2\over 2} {-} \beta_1 {3 q^2\over 16} {+} \beta_2 6 q^2\right),\,\,
\Gamma^2_{12}{\approx} {1\over \Mp}\left(\sqrt{3\over 2} {+} \beta_1 {\sqrt{3}\over 4\sqrt{2}} {+}\beta_2 2\sqrt{6}\right),\,\,
\Gamma^2_{22}{\approx}  {4\beta_2\over q^2}
\eea
where the metric to second order in $\beta_i$ takes the form
\bea
G_{11} &\approx& 1 - {5\beta_1\over 4} + \beta_2\left(-2 +  {11\beta_1\over 4} + \beta_2 4\right)\,,\quad  \quad G_{12} \approx M_P\beta_2\left( {\sqrt{3}\beta_1\over \sqrt{2}\,q^2 } +  {2\sqrt{6}\beta_2\over q^2}\right)\,, \\
G_{22} & \approx& M_P^2\beta_2\left({4\over 3 (q^2)^2} - {2\beta_1\over 3 (q^2)^2} + {16\beta_2\over 3 (q^2)^2}\right)~,
\eea
which leads to
\bea
V_{;11}&\approx& V_{11} - {3 q^2 \over 2 \Mp^2}V_2 + \beta_1\left(-{5\sqrt{3}\over 8\sqrt{2}\Mp} V_1
+ {3 q^2 \over 16\Mp^2}V_2\right) +\beta_2\left({\sqrt{6}\over \Mp} V_1 - {6 q^2\over \Mp^2} V_2\right)~, \cr
V_{;12} & \approx& V_{12} -{\sqrt{3\over 2}\over \Mp} V_2 -\beta_1 {\sqrt{3} \over 4\sqrt{2}\Mp} V_2 +
\beta_2\left({2\over q^2} V_1 -{2\sqrt{6}\over \Mp} V_2\right)~, \cr
V_{;22} &\approx& V_{22} +\beta_2\left({2\sqrt{2} \Mp\over \sqrt{3} (q^2)^2}V_1 -{4\over q^2}V_2\right)~,
\cr
\nabla_1 V_{;11} &\approx& V_{111} -{9 q^2 \over 2\Mp^2} V_{12} + \beta_1\left({15 \over 16\Mp^2}V_1 -{15\sqrt{3} \over 8\sqrt{2}\Mp} V_{11} + {9 q^2\over 16\Mp^2} V_{12} -{3\sqrt{3} q^2\over 16\sqrt{2}\Mp^3} V_2\right)   \cr
&& + \beta_2\left({6\over \Mp^2} V_1 + {3\sqrt{6}\over \Mp} V_{11} - {18 q^2\over \Mp^2} V_{12} - {6 \sqrt{6} q^2\over \Mp^3} V_2\right)~, \cr
\nabla_2 V_{;11} &\approx& V_{112} -{\sqrt{6}\over \Mp} V_{12} -{3\over 2\Mp^2} V_2 - {3 q^2\over 2\Mp^2} V_{22} +
\beta_1\left(-{9\sqrt{3}\over 8\sqrt{2}\Mp} V_{12} + {3\over 16\Mp^2} V_2 + {3 q^2\over 16\Mp^2} V_{22}\right) \cr
&&+ \beta_2\left({2\sqrt{6} \over q^2\Mp} V_1 + {4\over q^2} V_{11} - {3\sqrt{6}\over \Mp} V_{12} -{18\over\Mp^2} V_2 - {6 q^2\over \Mp^2} V_{22}\right)~, \cr
\nabla_1 V_{;12} &\approx& V_{112}-{\sqrt{6}\over\Mp} V_{12} -{3 q^2\over 2\Mp^2} V_{22} +\beta_1\left(-{9\sqrt{3}\over 8\sqrt{2}\Mp} V_{12} + {3\over 8\Mp^2} V_2 + {3q^2\over 16\Mp^2} V_{22}\right) \cr
&& + \beta_2\left({2\sqrt{6}\over q^2\Mp} V_1 + {4\over q^2} V_{11} - {3\sqrt{6}\over\Mp} V_{12} - {12\over\Mp^2}V_2 -{6 q^2\over\Mp^2} V_{22}\right)~, \cr
\nabla_2 V_{;12} &\approx& V_{122}-{\sqrt{6}\over\Mp} V_{22} - \beta_1 {\sqrt{3}\over 2\sqrt{2}\Mp} V_{22}
+\beta_2\left({2\over (q^2)^2}V_1 +{2\sqrt{2}\Mp\over \sqrt{3} (q^2)^2} V_{11} -{4\sqrt{6}\over q^2\Mp} V_2 - {4\sqrt{6}\over\Mp} V_{22}\right)~, \cr
\nabla_1 V_{;22} &\approx& V_{122}-{\sqrt{6}\over\Mp} V_{22} -\beta_1 {\sqrt{3}\over 2\sqrt{2}\Mp} V_{22}
+\beta_2\left({4\over (q^2)^2}V_1 +{2\sqrt{2}\Mp\over \sqrt{3} (q^2)^2} V_{11} -{4\sqrt{6}\over q^2\Mp} V_2 - {4\sqrt{6}\over\Mp} V_{22}\right)~, \cr
\nabla_2 V_{;22} &\approx& V_{222} +\beta_2\left({2\sqrt{6}\Mp\over (q^2)^2} V_{12} -{4\over (q^2)^2} V_2 - {12\over q^2} V_{22}\right)~.
\eea
where the regular partial derivatives are obtained by differentiating the potential with respect to $x$ and $y$ and then using the chain rule:
\begin{eqnarray}
V_{1} 
&=& \Mp^{-1}\,\Mp^4\hat\g x_0^2\,\sqrt{\frac{3}{2}}\Big(-2+4y_0\,\left(z_1-4z_2\right)-9z_3\Big)\cr
V_{11}
&=& \Mp^{-2} \,\Mp^4\hat\g x_0^2\,\Big(6 + 12y_0z_1\left(1-y_0\right) + 48y_0 z_2\left(2y_0-1\right) + \frac{81}{2}z_3\Big)\cr
V_{111} 
&=& \Mp^{-3}\,\Mp^4\hat\g x_0^2\,\frac{3}{2}\sqrt{\frac{3}{2}}\Big(-8 + 16y_0 z_1\left(1-3y_0+y_0^2\right)-64y_0 z_2\left(4y_0^2-6y_0+1\right) - 81z_3\Big)\cr
V_2
&=& \frac{8}{3} (q^2)^{-1}\,\Mp^4 \hat\g x_0^2\Big(z_1\left(y_0-1\right)- z_2\left(4 y_0-1\right)\Big)\cr
V_{22} 
&=& -\frac{8}{9} (q^2)^{-2}\,\Mp^4\hat\g x_0^2\,\Big(z_1\left(1-9y_0+4y_0^2\right) + z_2\left(1+20y_0-32y_0^2\right)\Big)\cr
V_{222} 
&=&  \frac{16}{27} (q^2)^{-3}\,\Mp^4\hat\g x_0^2\,\Big(z_1\left(1+11y_0-30y_0^2+8y_0^3\right) +2z_2\left(1+2y_0+72y_0^2-64y_0^3\right)\Big)\cr
V_{12}
&=& \frac{4}{3}(q^2)^{-1}\Mp^{-1}\,\Mp^4\hat\g x_0^2\,\left(-2\sqrt{6} y_0\right)\Big(z_1\left(y_0-2\right) + 2z_2\left(3-4y_0\right)\Big)\cr
V_{122} 
&=& \frac{8}{3}\sqrt{\frac{2}{3}}\,(q^2)^{-2}\,\Mp^{-1}\,\Mp^4\hat\g x_0^2\,y_0\Big(z_1\left(10-17y_0+4y_0^2\right)-2 z_2\left(9-52y_0+32y_0^2\right)\Big)\cr
V_{112}
&=& 16(q^2)^{-1}\,\Mp^{-2}\,\Mp^4\hat\g x_0^2\,y_0\,\Big(z_1\left(2-4y_0+y_0^2\right) -2z_2\left(3-14y_0+8y_0^2\right)\Big)\,.
\label{d-new}
\end{eqnarray}

\section{Computation of the observables}
\label{sec:obser}

The power spectrum of adiabatic perturbations is given by \cite{lalak2007}:
\begin{equation}
{\cal P} \approx \left({H^2\over 2\pi \dot\sigma}\right)^2={H^4\over 2 (2\pi)^2 K}\,,\quad {\rm where}\quad \dot \sigma^2 = G_{ij}\dot q^i\dot q^j=2K~.
\end{equation}
Using the slow roll equations, and to lowest order in $\beta_i$ this is (\ref{slowrollpower})
\begin{equation}
{\cal P} \approx {1 \over 12 \pi^2} {V^3 \over M_P^6 V_1^2} =
 \frac{2}{3\pi^2}\,\frac{\hat\g\,x_0^2}{16 \, \hat{\epsilon}}\,\left(1 + 3  z_3 + 4  z_2 - 2   z_1\right)\,.
 \label{powerz}
\end{equation}
where, in the last expression, we used (\ref{tecpot}) and the expressions in Appendix A to write ${\cal P}$ in terms of the effective parameters  $z_i$.    Similarly, the square root of  tensor to scalar ratio (\ref{sqrtr}) can be expressed as
\be
\sqrt{r} \approx \sqrt{8} M_P {V_1 \over V}
= s\,2\sqrt{3}\,\frac{2 - 4 y_0(z_1 -  z_2) + 9 z_3}{1 + 3  z_3 + 4  z_2 - 2   z_1}\,,
\ee
where $s$ is the sign of the slope of the potential $s=\text{sign}(V_1)$\,.

The spectral index $n_s$ is defined as,
\begin{equation}
n_s{-}1 = \frac{d\log{\cal P}}{d\log k} \approx H^{-1} \frac{\dot {\cal P}}{{\cal P}}\,,
\end{equation}
where we traded the derivative with respect to the wave number $k$ with a time derivative because close to the Hubble horizon crossing surface the following set of approximations hold \cite{stewart} :
\begin{align*}
\frac{d}{d \log\,k} \approx &\frac{d}{d \log\,(aH)} = \frac{d \log a}{d \log\, (aH)}\,\frac{d}{d \log\,a} 
= \frac{d\log a}{dt}\, \frac{1}{\frac{d \log\, (aH)}{dt}}\,  
\frac{d}{\frac{d \log\,a}{dt}}\,\frac{d}{dt} \\
= & \,H\,\frac{1}{H + \frac{\dot H}{H}}\, H^{-1}\,\frac{d}{dt}\approx H^{-1}\,\frac{d}{dt}\,,
\end{align*}
to lowest order in the slow-roll parameters, since $\dot H/H^2 = -\hat \epsilon \ll1$. Since
\begin{equation*}
  \dot{{\cal P}} = {\cal P}\,\left(4\frac{\dot{H}}{H} - \frac{\dot{K}}{K}\right)\,, \quad 
  \frac{\dot{K}}{H\,K} = -2\frac{\dot{H}}{H^2} - \frac{1}{3H^2\,K}\,\dot{q}^i\,\dot{q}^j\,V_{;ij}\,,
\end{equation*}
and since $\dot{H} = -K/M_P^2$ from the equations of motion, we derive the final expression 
\be
n_s {-} 1 \approx - 6 \hat{\epsilon} + 2 \hat{\eta}
\ee
for the spectral index reported in \eqref{ns}.    Using the expressions (\ref{firstslowroll12},\ref{etahat12}) for the slow roll parameters and the results in Appendix A, the spectral index can be written as
\bea
  n_s {-}1 &=& -{9\over 2}\left({2-4y_0(z_1-4z_2)+9z_3\over1-2 z_1 + 4 z_2 + 3 z_3}\right) 
  + {12 +24 y_0 z_1(1-y_0) + 96 y_0 z_2 (2 y_0 - 1) + 81 z_3\over 1-2 z_1 + 4 z_2 + 3 z_3}
  \nonumber \\
 &&-
 96 y_0^2 {\left((y_0-2) z_1 + (6-8y_0)z_2\right)^2\over \left(z_1(1-9 y_0 + 4y_0^2) + z_2(1+20 y_0 - 32 y_0^2\right)\left(1-2 z_1 + 4 z_2 + 3 z_3\right)}
\,
\eea
in terms of the microscopic parameters.

The computation of the running $\alpha_s$ involves :
\begin{equation*}
  \alpha_s \approx \frac{6}{H}\,\frac{d\dot{H}/H^2}{dt} + \frac{2}{H}\,\frac{d\hat\eta}{dt}\,.
\end{equation*}
Since $\dot{H}\approx -K$ and we already computed the rate of variation of the kinetic energy density $K$ during slow roll, the first term is easy to compute and it equals :
\begin{equation}
  \frac{6}{H}\,\frac{d\dot{H}/H^2}{dt} \approx -24\,\hat\epsilon^2 + 12\,\hat\epsilon\,\hat\eta\,.
 \label{runc1}
\end{equation}
The second term has different contributions :
\begin{equation*}
  \frac{2}{H}\,\frac{d\hat\eta}{dt} \approx 4\hat\epsilon\,\hat\eta - 2\frac{\dot{K}}{H\,K}\,\hat\eta + \frac{1}{3H^3\,K}\,V_{;ijk}\dot{q}^i\,\dot{q}^j\,\dot{q}^k + \frac{2}{3H^3K}\,V_{;ij}\,\dot{q}^i\,\frac{d\dot{q}^j}{dt}\,.
\end{equation*}
Using the identity :
\begin{equation*}
  \frac{d\dot{q}^j}{dt} \approx -\frac{\dot{H}}{H}\,\dot{q}^j - \frac{1}{3H}\,\dot{q}^k\,G^{jm}\,V_{;mk}\,,
\end{equation*}
we find that
\begin{equation}
  \frac{2}{H}\,\frac{d\hat\eta}{dt} \approx 4\,\hat\epsilon\,\hat\eta - 2\,\hat\psi + 4\left(\hat\eta^2-\widehat{(\eta^2)}\right)\,.
  \label{runc2}
\end{equation}
Adding \eqref{runc1} and \eqref{runc2}, we obtain the final expression for the running of the spectral index in \eqref{run}.

\paragraph{Cancellation of $\hat\eta^2-\widehat{(\eta^2)}$ :} Let us calculate this term for the two field models considered in this paper. The expression for $\hat\eta$ derived from its definition and valid during the slow roll regime analysed here is: 
\begin{equation*}
  \hat{\eta} = M_P^2 {V_{11}V_{22} -(V_{12})^2\over V V_{22}} + \eO(\beta_i)\,.
\end{equation*}
On the other hand,
\begin{eqnarray*}
  \widehat{(\eta^2)} &\approx& {M_P^4\over  V^2 (\dot q^1)^2}\left(\dot q^j\dot q^kV_{;1k} V_{1j} + \dot q^j\dot q^kV_{;2k} V_{2j}G^{22}\right)(1+ \eO(\beta_i)) \\
  &\approx& M_P^4\left({V_{11}V_{22} -(V_{12})^2\over V V_{22}}\right)^2+ \eO(\beta_i)\,.
\end{eqnarray*}
In the first approximation we have used the standard slow roll approximation for relating $H^2$ and $V$ \eqref{slowroll1} and also expanded the kinetic energy to leading order in $\beta_i$ \eqref{kineticexpansion}. We then use the expressions for the relationship between the $\dot q^i$'s \eqref{slow-velocity} due to the slow-roll constraints \eqref{slowroll2a} and the result for the covariant derivatives given in Appendix~A.
Thus it follows that 
\begin{equation}
  \hat\eta^2-\widehat{(\eta^2)} \approx \eO(\beta_2)\,.
\end{equation}
In particular, all the potentially large contributions being proportional to $1/\beta_2$ cancel out, due to our choice of initial conditions, and giving further evidence that this is a slow roll regime.

\section{Parameters as functions of initial conditions and the observables} 

Using the slow-roll constraint we found that $z_2$ was determined in terms of $z_1$ (\ref{constraint1}) as
\begin{equation}
  z_2 = \frac{y_0-1}{4y_0-1}\,z_1 + \eO(\beta_0\,\sqrt{r})\,.
\end{equation}
Then we can invert the expressions for $n_s {-}1$ and $r$ obtained in Appendix B to obtain 
\begin{eqnarray}
  z_1 &=& \frac{c}{d}\,, \\
  c &=& (4y_0-1)\,(1+3y_0-6y_0^2+8y_0^3)\,\left(288(17+n_s)\,y_0 + s\,3\sqrt{3}r^{3/2}\,(1+2y_0) \right. \nonumber \\
  & & \left. + s\,8\sqrt{3}\sqrt{r}\left(17+n_s+2y_0(62+n_s)\right) + 12 r(5+19y_0)\right)\,, \nonumber \\
  d &=& 6\,(12\,\sqrt{3}\,y_0 + s\,\sqrt{r}\,(1+2y_0))\,\left(8\sqrt{3} n_s\,(1+3y_0-6y_0^2+8y_0^3) \right. \nonumber \\
  & & \left. +3\sqrt{3} r(1+3 y_0 - 6 y_0^2 + 8 y_0^3)+s\,12\sqrt{r}(3+19 y_0-40 y_0^2+48 y_0^3) \right. \nonumber \\
  & & \left. +8\sqrt{3}(-1+87y_0-192 y_0^2 + 208 y_0^3)\right)\,.
\end{eqnarray}
and
\begin{eqnarray}
  z_3 &=& \frac{a}{b}\,, \\
  a &=& 2\,(y_0-1)\,\left(3\sqrt{3}r\,(1+3y_0-6y_0^2+8y_0^3)+ s\,24\,\sqrt{r}\,(1+9y_0-21y_0^2+20y_0^3) \right.\nonumber \\
  && \left. + 8\,\sqrt{3}(-1+69y_0-174y_0^2+136y_0^3+n_s\,(1+3y_0-6y_0^2+8y_0^3))\right)\,, \nonumber \\
  b &=& 9\,\left(3\sqrt{3}r\,(1+3y_0-6y_0^2+8y_0^3) +s\,12\,\sqrt{r}\,(3+19 y_0-40 y_0^2 + 48 y_0^3) \right. \nonumber \\
  & & \left. + 8\sqrt{3}\,(-1+87y_0-192y_0^2+208y_0^3+n_s\,(1+3y_0-6y_0^2+8y_0^3))\right)\,.
\end{eqnarray}
To  lowest order in a power expansion in $r$ :
\begin{eqnarray}
  z_1 & \approx &  \frac{(17+n_s)\,(-1+y_0 + 18 y_0^2 - 32 y_0^3 + 32 y_0^4)}{6(-1+ 87 y_0 - 192 y_0^2 + 208 y_0^3 + n_s(1+ 3 y_0 - 6 y_0^2 + 8 y_0^3)} + \eO(\sqrt{r})\,, \cr
  z_3 & \approx & \frac{2\,(y_0-1)\,(-1+ 69 y_0- 174 y_0^2 + 136 y_0^3 +  n_s\,(1+ 3 y_0 - 6 y_0^2 + 8 y_0^3)}{9 (-1+ 87 y_0 - 192 y_0^2 + 208 y_0^3 + n_s\,(1+ 3 y_0 - 6 y_0^2 + 8 y_0^3)} + \eO(\sqrt{r})\,.
\end{eqnarray}


The value of the running $\alpha_s$ in the same approximation is given by
\begin{multline}
  \alpha_s \approx \frac{\sqrt{3}\,s\,\sqrt{r}}{4\,(1-y_0+4y_0^2)(1+3y_0-6y_0^2+8y_0^3)^2}\cdot \\
  \left[-5+315 y_0 -1437 y_0^2 + 2889 y_0^3 +1886 y_0^4 - 14272 y_0^5 + 23360 y_0^6 - 22784 y_0^7 + 11776 y_0^8 \right. \\
  \left. + n_s\left(5+ 45 y_0 - 111 y_0^2 + 207 y_0^3 + 598 y_0^4 - 2216 y_0^5 + 4000 y_0^6 - 3712 y_0^7 + 2048 y_0^8\right)\right] \\
  +\eO(r)\,,
\end{multline}
whereas the higher order corrections in $r$ can be safely neglected whenever $y_0 \in (5,\,\infty)$.

Finally, if both  $r$ and $n_s {-} 1$ vanish,  the $z_i$ must be related to the initial conditions as:
\begin{eqnarray}
z_1^0 &=& {(4y_0 - 1) ( 1 + 3 y_0 - 6 y_0^2 + 8 y_0^3) \over 6 y_0 (5 - 11 y_0 +12 y_0^2)}
\label{z1limit}\\
z_2^0 &=& {(y_0 -1) ( 1 + 3 y_0 - 6 y_0^2 + 8 y_0^3) \over 6 y_0 (5 - 11 y_0 + 12 y_0^2)}
\label{z2limit} \\
z_3^0 &=& {4 (y_0 - 1) (2 - 5 y_0 + 4 y_0^2) \over 9 (5- 11 y_0 + 12 y_0^2}
\label{z3limit}
\end{eqnarray}

\end{document}